\newcommand\ket [1] {|#1 \rangle }
\newcommand\bra [1] {\langle #1 |}

\newcommand{\av}[1]{\langle #1\rangle}
\newcommand{\bb}[1]{\mathbf{#1}}

\documentclass{ws-ijmpb}
\usepackage[english]{babel}
\usepackage{amsmath}
\usepackage[colorlinks,bookmarks=true,citecolor=blue,linkcolor=red,urlcolor=blue]{hyperref}

\def\onlinecite[#1]{\cite[left=,right=][#1]}

\begin{document}
\markboth{Emil J. Bergholtz and Zhao Liu}
{Topological Flat Band Models and Fractional Chern Insulators}

%
\catchline{}{}{}{}{}

\title{Topological Flat Band Models and Fractional Chern Insulators}

\author{Emil J. Bergholtz}

\address{Dahlem Center for Complex Quantum Systems,\\ Institut f\"ur Theoretische Physik, Freie Universit\"at Berlin,\\ Arnimallee 14, 14195 Berlin, Germany
\\
ejb@physik.fu-berlin.de}

\author{Zhao Liu}

\address{Beijing Computational Science Research Center,\\ Beijing, 100084, China
\\
zliu@csrc.ac.cn}

\maketitle

\begin{abstract} Topological insulators and their intriguing edge states can be understood in a single-particle picture and can as such be exhaustively classified. Interactions significantly complicate this picture and can lead to entirely new insulating phases, with an altogether much richer and less explored phenomenology. Most saliently, lattice generalizations of fractional quantum Hall states, dubbed fractional Chern insulators, have recently been predicted to be stabilized by interactions within nearly dispersionless bands with non-zero Chern number, $C$. Contrary to their continuum analogues, these states do not require an external magnetic field and may potentially persist even at room temperature, which make these systems very attractive for possible applications such as topological quantum computation. This review recapitulates the basics of tight-binding models hosting nearly flat bands with non-trivial topology, $C\neq 0$, and summarizes the present understanding of interactions and strongly correlated phases within these bands. Emphasis is made on microscopic models, highlighting the analogy with continuum Landau level physics, as well as qualitatively new, lattice specific, aspects including Berry curvature fluctuations, competing instabilities as well as novel collective states of matter emerging in bands with $|C|>1$. Possible experimental realizations, including oxide interfaces and cold atom implementations as well as generalizations to flat bands characterized by other topological invariants are also discussed.
\end{abstract}


\section{Introduction and scope}
The discovery of the quantum Hall (QH) effect,\cite{vonk,tsui} manifested by a remarkably precise quantization of the transverse conductance in effectively two-dimensional electron systems in presence of strong perpendicular magnetic fields, has had profound implications for the understanding of matter. The integer QH effect,\cite{vonk} is the first example of a topological insulator,\cite{toprev1,toprev2} and can as such be understood in a single particle framework:\cite{it1,it2} charged particles in a magnetic field form Landau levels with energy splitting that is proportional to the strength of the magnetic field, and when an integer number of Landau levels are filled a band insulator with a bulk gap forms. In contrast to ordinary band insulators, filled Landau levels come with gapless chiral edge states that each carries a quantum of conductance, $e^2/h$. Mathematically, the number of edge states is given by the value of a topological invariant, namely the Chern number, that can only assume integer values similar to a winding number. The integer nature of the Chern number is what makes the edge states, and hence the quantization of the (off-diagonal) conductivity, so remarkably robust---they are "topologically protected" and as such entirely insensitive to disorder as long as the bulk gap does not close.\cite{tknn,avron,qniu} In fact this quantization is so precise that it has lead to a new laboratory definition of resistance in terms of the von Klitzing constant, $R_K=h/e^2=25812.807557(18)\Omega$.\cite{resdef}

The fractional quantum Hall (FQH) effect,\cite{tsui} observed when the number of filled Landau levels is a fraction $\nu=p/q$, is, albeit showing strikingly similar transport features, of a very different origin:\cite{Laughlin83} it is entirely stabilized by interactions within the exponentially large manifold of states which are (exactly) degenerate at the single particle level due to dispersionless nature of Landau levels. In contrast to the topological insulators, the FQH states fall within the realm of topological order\cite{toporder,wenbook} and are as such characterized by non-trivial ground state degeneracies depending on the genus of the manifold on which they live,\cite{wenniu} long-range entanglement\cite{kitaevpreskill,levinwen} and fractionalized excitations.\cite{Laughlin83,mr} Motivated by the fundamental interest in finding new types of particles, as well as by the search for robust quantum computational devices,\cite{topocomp} much of the recent interest in FQH physics has focused on phases with non-Abelian excitations.\cite{mr} In contrast to fermions, bosons, or Abelian anyons\cite{fstat} for that matter, the non-Abelian quasiparticles have the property that the wave function in general becomes linearly independent from the starting state when the positions of two (or more) of them are adiabatically interchanged (braided). This potentially provides excellent degrees of freedom to store information (the quantum bit) that is immune to local perturbations such as disorder and can only be altered by non-local braiding operations of the excitations.\cite{topocomp} Although these theoretical ideas are well developed and sophisticated, topological quantum computation remains a rather remote dream to this date. Despite old ideas in the context of high-energy physics,\cite{majoranareturns} and more recently impressive solid state experiments, in the quantum Hall regime\cite{willett} as well as in quantum wires,\cite{mourik} there is so-far no firmly established realization of non-Abelian (quasi)particles in nature. Moreover, the large scale manipulation needed for actually performing quantum computation is posing an enormous technological challenge. In the conventional quantum Hall setup using semiconductor heterostructures, two main limitations are the need for a very strong magnetic field $B\sim 10$ Tesla, and the fact that the gap nevertheless remains very small $\Delta E \lesssim 1$ Kelvin. Another key issue is the need for ultra clean samples with extremely high mobility, especially for more fragile FQH phases, including the non-Abelian ones.

A conceptually important step towards high-temperature topological phenomena was provided by Haldane in a seminal work, published already in 1988, where he explicitly showed that an integer quantum Hall effect can in fact appear also without an external magnetic field, by constructing a simple tight-binding model on the honeycomb lattice.\cite{haldanemodel} In this model, which was the first lattice construction of a Chern insulator, time-reversal symmetry is broken by a magnetic field of zero average flux, which can, for instance, be emulated by spin-orbit coupling. In fact, the Haldane model also underlies the recent developments in topological insulators---the lattice version of the archetypical Kane-Mele model\cite{kanemele,kanemele2} is essentially built by two (time-reversed) copies of the Haldane model.

On a lattice the effect of a magnetic flux through a plaquette is actually physically indistinguishable when an integer number of flux quanta, $\Phi_0=h/(2e)$ is added. This observation reveals a close link between the Haldane model and the problem of charged particle hopping on a (square) lattice in presence of a perpendicular magnetic field which was studied already in 1969 by Hofstadter\cite{Hofstadter} and is known to exhibit an intriguing fractal spectrum, the famous "Hofstadter Butterfly", as a function of the flux per plaquette. In the limit of small flux this recovers Landau levels in the continuum.

The recent interest in fractional Chern insulators (FCIs) was ignited\cite{viewpoint1} by the insight that the topological bands can be made flat for suitable short-range tight-binding parameters,\cite{kapit,chernins1,chernins2,chernins3} thus greatly increasing the effect of interactions. Early numerical works\cite{chernins3,cherninsnum1,cherninsnum2} indeed confirmed the existence of electronic FCI analogues of the Laughlin state.\cite{Laughlin83} [See Refs.\cite{kapit,bosons} for corresponding results for bosons.] These results opened a number of intriguing possibilities. First, the gap, which is mainly controlled by the Coulomb interaction, $\Delta E\sim e^2/(\epsilon \ell)$, may be greatly increased compared to the continuum as it is stabilized by interactions on the lattice scale, rather than on the order of the magnetic length which is typically two orders of magnitude larger.
In fact, a naive estimate does not rule out FCIs at room-temperature.\cite{chernins3}
Second, there is no need for a strong external magnetic field. The two effects of the magnetic field in the continuum, namely the breaking of time reversal symmetry and the formation of flat bands, can be replaced by a suitable combination of e.g., spin-orbit coupling and ferromagnetism. Third, lattice systems can harbor qualitatively new phases of matter. An exciting example thereof is the existence of relatively realistic flat band models with Chern number larger than one, $|C|>1$,\cite{c2,max,dassarma} some of which have been confirmed to host a plethora of new FCIs markedly beyond the FQH paradigm.\cite{ChernN,ChernTwo,ChernN2} A particularly intriguing possibility is that lattice defects in these systems may act as worm-hole-like objects, suitably dubbed "genons", that change the geometry of space and obey non-Abelian exchange statistics.\cite{disloc}

Along with the new possibilities the topological flat band systems also pose a number of important new theoretical challenges that arise due to the combined effects of interactions, band topology and the underlying lattice. One of the ways in which this is manifested is through a varying Berry curvature which is a necessary complication compared to the continuum case. A related general lattice feature is the lack of translational invariance (in reciprocal space), and a resulting absence of particle hole symmetry in a band and the emergence of qualitatively new competing phases.\cite{andreas}

The search for experimental realizations of FCIs is arguably the most pressing issue in the field. In this context, it is encouraging to consider recent history. The success story of topological insulators nicely illustrates that wild ideas,\cite{kanemele,kanemele2} followed by detailed theoretical predictions\cite{BHZ} can lead indeed to ground-breaking experiments.\cite{topoexp} Moreover, in analogy with the development in the quantum Hall context it is very encouraging to note that, very recently, the first experimental realization of a Chern insulator was reported in a magnetic topological insulator\cite{Chernexp} and that signatures of the Hofstadter butterfly have been observed in graphene superlattices\cite{hofstadtergraphene1,hofstadtergraphene2} as well as in a cold atom experiment\cite{hofstadtercold}.
Indeed, there is a rapidly growing list of more or less detailed suggestions for realizing FCIs ranging from solid state materials such as oxide interfaces\cite{digital,ifw2} to cold atom realizations\cite{DipolarTFB2} and so-called optical flux lattices.\cite{opticalflux}

Although the recent interest in these systems has uncovered plenty of new physics, there is a rich history of earlier works that are worth mentioning. In particular, it was noticed almost twenty years ago that a periodic potential in addition to the Landau level structure can lead to qualitatively new insulating phases, most saliently with an unexpected value of the quantized conductance.\cite{kolread} Moreover, various FQH states have been found in the context of the Hofstadter model, first in the weak field limit\cite{sorensen} (i.e. close to the continuum) and later also when the effect of the lattice is considerable\footnote{The early evidence for lattice FQH states was not quite as unambiguous as in later works due to the limitations to very small systems sizes and, more importantly, since the bands of the Hofstadter model are not very flat.}.\cite{palmer,gunnar} Yet another line of precursor studies stems from the idea of chiral spin liquids\cite{kalmeyerlaughlin} in the context of frustrated magnetism.

The present review aims to fill two main purposes. First, it provides a detailed introduction to many of the basic concepts that should serve as a useful reference for anyone entering the field. Second, it provides a snapshot of the state of the art in the field and a collection of references to the relevant original works. To illustrate the general points that we want to make, we include a number of explicit examples with previously unpublished data and our interpretations thereof. In addition to this we highlight a number of directions that we feel deserve future attention. Recently another introduction to the field, albeit with a different focus, appeared.\cite{otherreview} The excellent discussion in Ref.\cite{otherreview} regarding the algebra of band projected density operators is recommended as a complementary reading as this topic is only briefly touched upon here. Another omission in our work is parton constructions and effective field theory approaches, partly because of space limitations, but also because at least some of these approaches are associated with considerable unsolved technical as well as conceptual issues when applied to the lattice systems. Nevertheless, some of this stream of works are interesting and we direct the interested readers to a selection of original publications.\cite{nthroot,cherncf2} Also worth mentioning is the recent review in Ref.\cite{hohenadler} that surveys various recent studies on effects of electron correlations in topological insulators, mainly focusing on the integer regime and on time-reversal invariant systems.

The rest of this review is organized as follows. In Section \ref{nonint} we give a rather detailed account of topological flat band models, with focus on Chern bands, including comments on technical details and subtleties. In Section \ref{bandproj}, we introduce interactions and band projections. Section \ref{c1} reviews the present understanding of interaction phases in $|C|=1$ bands, highlighting both the similarities, e.g. through adiabatic continuity, as well as differences including new competing phases, compared to the conventional Landau level phenomena in the continuum. In Section \ref{Cg1} we move on to flat band physics qualitatively beyond the FQH paradigm by reviewing very recent works on models with $|C|>1$. In Section \ref{exp} we discuss some of the most promising ideas that have been put forward for the experimental realization of FCIs. Finally, in Section \ref{discussion}, we close by a discussion of a number of intriguing future directions in this field.

\section{Tight-binding models, Berry curvature, and flat bands}\label{nonint}

Consider a translation invariant quadratic hopping Hamiltonian,
\begin{equation}
H_0=\sum_{n,m} t^{ab}_{nm}c^\dagger_{n,a}c_{m,b} \ , \label{hop}
\end{equation}
where $a,b=1,\ldots,N_c$ label the states in the unit cell, and $n,m=1,\ldots,N_s$ label the sites on the Bravais lattice (at positions ${\bb R}_n$ and ${\bb R}_m$). With $c^\dagger_{{\bb k},a}=\frac 1 {\sqrt{N_s}}\sum_n e^{i{\bb k}\cdot {\bb R}_n}c^\dagger_{n,a}$, one finds
\begin{equation}
H_0=\sum_{a,b,{\bb k}} \mathcal H_\bb k^{ab}c^\dagger_{{\bb k},a}c_{{\bb k},b}
\ , \label{bloch}
\end{equation}
where ${\bb k}=(k_1,k_2)$ is the single-particle momentum restricted to the first Brillioun zone (BZ) and $\mathcal H_\bb k^{ab}\equiv\frac 1 {N_s}\sum_{n,m} t^{ab}_{nm}e^{-i{\bb k}\cdot ({\bb R}_n-{\bb R}_m)}=\sum_{n} t^{ab}_{n1}e^{-i{\bb k}\cdot ({\bb R}_n-{\bb R}_1)}$ [${\bb R}_1$ can be set to zero for convenience]. In reciprocal space, the single-particle (Bloch) Hamiltonian, $\mathcal H_\bb k$, is diagonalized,
\begin{equation} \mathcal \sum_{a,b} \mathcal H_\bb k^{ab}c^\dagger_{{\bb k},a}c_{{\bb k},b}\ket{\psi_s(\mathbf k)}=E_s(\bb k)\ket{\psi_s(\mathbf k)}\ ,\label{diag}\end{equation}
for each $\bb k$ separately, by the states $\ket{\psi_s(\mathbf k)}=\sum_b\psi^b_s(\bb k)c^\dagger_{{\bb k},b}\ket{0}$. The energies $E_s(\bb k), s=1,\ldots,N_c,$ are the eigenvalues of the matrix $\mathcal H_\bb k$ and constitute the band structure of the model (\ref{hop}) with $\bb k\in {\rm BZ}$.

To characterize the topological properties of a given band it is useful to calculate the Chern number,
\begin{equation} C=\frac 1 {2\pi}\int_{\rm{BZ}} F^s_{12}({\mathbf k})\mathrm{d}^2{\mathbf k}\in \mathbb Z\ ,\label{chernnumber}\end{equation} which is an integer valued quantity. In Eq.~(\ref{chernnumber}), $C$ is defined for an isolated band, $s$, in terms of the wave functions $\ket{\psi_s(\mathbf k)}$, via the Berry curvature,
\begin{equation} F^s_{\alpha\beta}({\mathbf k})=\partial_{k_\alpha}A^s_\beta({\mathbf k})-\partial_{k_\beta}A^s_\alpha({\mathbf k})\ ,\label{berrycurvature}\end{equation} where
\begin{equation} A^s_\beta({\mathbf k})=-i\bra{\tilde\psi_s(\mathbf k)}\partial_{k_\beta}\ket{\tilde\psi_s(\mathbf k)}\label{berryconnection}\end{equation} is the Berry connection, $\alpha,\beta=1,2$ and $\ket{\tilde\psi_s(\mathbf k)}\equiv e^{-i\mathbf k\cdot \hat{\bb r}}\ket{\psi_s(\mathbf k)}=\frac{1}{\sqrt{N_s}}\sum_{n,b} e^{-i\mathbf k\cdot \delta\bb r_b}\psi^b_s(\bb k)c^\dagger_{n,b}|0\rangle$. Here $\delta\bb r_b$ is the relative position with respect to the center of the unit cell (i.e., the sites of the lattice are located at ${\bb R}_n+\delta\bb r_b$). There is a subtle, often overlooked, difference between using $\ket{\tilde\psi_s(\mathbf k)}$ and $\ket{\psi_s(\mathbf k)}$ in the Berry connection: while both choices give the same Chern number, the resulting Berry curvature is not identical. In the case of the Hofstadter butterfly\cite{Hofstadter} only the former gives a well behaved weak field continuum limit (constant Berry curvature). Thus the details of the embedding of the lattice model and its orbitals in real-space have observable effects as it influences the Berry curvature distribution.

Eqs.~(\ref{berrycurvature},\ref{berryconnection}) highlight the fact that the (gauge independent) Berry curvature and the (gauge dependent) Berry connection\footnote{Note that a gauge transform $\ket{\tilde\psi_s(\mathbf k)}\rightarrow e^{i\phi(\bb k)}\ket{\tilde\psi_s(\mathbf k)}$ changes the Berry connection (\ref{berryconnection}): $A^s_\beta({\mathbf k})\rightarrow  A^s_\beta({\mathbf k})+\partial_{k_\beta}\phi(\bb k)$, while the Berry curvature (\ref{berrycurvature}) remains unchanged as long as $\phi(\bb k)$ is differentiable.} are solely dependent on the eigenstates of the band (and their embedding in real-space), but are in principle independent of the band energy structure.
It is important to note that some care is needed when evaluating expressions including derivatives of the eigenstates, as in Eqs.~(\ref{berrycurvature},\ref{berryconnection}), since they implicitly assume a consistent (smooth) choice of gauge throughout the Brillioun zone. In practical calculations it is therefore preferable to use a formula where the energy eigenvalues of the model nevertheless show up:\cite{Berry}
\begin{eqnarray}
F^s_{\alpha\beta}({\mathbf k})&=&\sum_{s'\neq s}\frac{\bra{\tilde\psi_{s}(\mathbf k)}\frac{\partial \mathcal H_\bb k}{\partial{k_\alpha}}\ket{\tilde\psi_{s'}(\mathbf k)}\bra{\tilde\psi_{s'}(\mathbf k)}\frac{\partial \mathcal H_\bb k}{\partial{k_\beta}}\ket{\tilde\psi_s(\mathbf k)}-(\alpha\leftrightarrow \beta)}{[E_s(\bb k)-E_{s'}(\bb k)]^2}\nonumber \\
&=& 2 \sum_{s'\neq s}{\rm Im}\left \{\frac{\bra{\tilde\psi_{s}(\mathbf k)}\frac{\partial \mathcal H_\bb k}{\partial{k_\alpha}}\ket{\tilde\psi_{s'}(\mathbf k)}\bra{\tilde\psi_{s'}(\mathbf k)}\frac{\partial \mathcal H_\bb k}{\partial{k_\beta}}\ket{\tilde\psi_s(\mathbf k)}}{[E_s(\bb k)-E_{s'}(\bb k)]^2}\right \}
\ .\label{berrycurvature2}\end{eqnarray}
Eq.~(\ref{berrycurvature2}) can be derived by noting that $\bra{\tilde\psi_{s'}(\mathbf k)}\partial_{k_\beta}\ket{\tilde\psi_s(\mathbf k)}[E_s(\bb k)-E_{s'}(\bb k)]=\bra{\tilde\psi_{s'}(\mathbf k)}\frac{\partial \mathcal H_\bb k}{\partial{k_\beta}}\ket{\tilde\psi_s(\mathbf k)}$, for $s\neq s'$.

Physically, the Chern number counts the number of current carrying chiral edge states, and comes with a sign which indicates the direction of propagation of the chiral modes. Consequently, the Chern number is in direct proportion to the quantized Hall conductivity of a filled band,\cite{tknn} \begin{equation}\sigma_H=C\frac{e^2}{h}\ . \label{filledhc}\end{equation} When several bands are filled the Hall conductivity is the sum of the individual band contributions, each given by Eq.~(\ref{filledhc}). For a gapped many-body state at fractional band filling, the Hall conductance is not generally quantized. However, an intuitive and appealing expression for the Hall conductivity was provided in Ref.\cite{hallcond} as
 \begin{equation}\sigma_H=\frac 1 {2\pi}\int_{\rm{BZ}} F^s_{12}({\mathbf k})\av{n_\mathbf k}\mathrm{d}^2{\mathbf k}, \label{manybodyconductance}\end{equation}
where $\av{n_\mathbf k}$ is the occupation number averaged over (quasi-) degenerate ground states and is as such containing information about the correlations in the system\footnote{Reliable finite size results can be achieved by averaging over the boundary conditions, i.e. inserting flux through the handles of the torus on which the system is defined.}. Although interesting counterexamples exist,\cite{kolread,cherncf2,hallcond} Eq.~(\ref{manybodyconductance}) typically gives $\sigma_H=C\nu\frac{e^2}{h}$ for an incompressible state at fractional band filling, $\nu$.

Topologically non-trivial bands, with $C\neq 0$, can appear when the hopping parameters, $t^{ab}_{nm}$, are allowed to assume complex values which naturally arises in a number of systems including spin-orbit coupled materials and systems with effective gauge fields. It has also been argued (at the mean-field level) that effective complex hopping parameters can occur spontaneously due to strong frustrated interactions.\cite{topomott}

It is crucial to realize that the local Berry curvature and the (energy) dispersion are fundamentally independent. To see this it is instructive to consider a non-degenerate band, with energy $E_s(\bb k)\neq 0$, without any touching points with other bands\footnote{Note that the condition $E_s(\bb k)\neq 0$ can always be met by adding an appropriate constant to the Hamiltonian without changing the eigenstates.}. Now, a flat band model can be trivially constructed by the replacement
\begin{equation}\mathcal H_\bb k \rightarrow \mathcal H_\bb k ^{{\rm flat}}=\mathcal H_\bb k/ E_s(\bb k).\label{flattening}\end{equation}
While the dispersion of the band corresponding to $\mathcal H_\bb k ^{{\rm flat}}$ is entirely flat, the eigenstate $\ket{\psi_s(\mathbf k)}$ and thus the Berry curvature remain unaltered [cf. Eqs.~(\ref{berrycurvature},\ref{berryconnection}) and note that $E_s(\bb k)\in \mathbb R$]. It is not possible to have an entirely flat Berry curvature in any lattice model except in the limit of an infinitely large unit cell---in the latter limit one can obtain the continuum Landau levels which have asymptotically flat energy dispersion and Berry curvature.

In real-space, the flattened Hamiltonian,
\begin{equation}H_0^{{\rm flat}}=\sum_{a,b,m,n}t^{ab,{\rm flat}}_{nm}c^\dagger_{n,a}c_{m,b} \ ; \ \ \ \ \  t^{ab,{\rm flat}}_{nm}\equiv\frac{1}{N_s}\sum_{\bb k} \frac{\mathcal H^{ab}_\bb k}{E_s(\bb k)}e^{i{\bb k}\cdot ({\bb R}_n-{\bb R}_m)}\ ,\label{flatreal}\end{equation}
includes arbitrary long-range processes even if the original Hamiltonian (\ref{hop}) is short-range. What makes flat band models more than a curiosity is that the hopping amplitudes (asymptotically) decay exponentially and that keeping only a few short-range terms already often gives almost flat bands\footnote{A model where exponentially decaying hopping elements are explicitly written out was constructed in Ref.\cite{kapit}.}.

To quantify how good a flat band model is, it is useful to define the flatness ratio, $F=\Delta/W$, in terms of the bandwidth, $W=\max\limits_{\bb{k}, \bb{k}' \in \rm{BZ}}[E_{s}(\bb{k})-E_{s}(\bb{k}')]$, and the energy gap, $\Delta=\min\limits_{\bb{k}, \bb{k}' \in \rm{BZ}}[E_{s}(\bb{k})-E_{s-1}(\bb{k}'), E_{s+1}(\bb{k}')-E_{s}(\bb{k})]$. While only $F>0$ is needed to have an insulator at the non-interacting level, a separation of energy scales at large $F\gg 1$ implies that there is a window of interaction strength, $V$, such that
\begin{equation}\Delta \ {\rm (band\ gap)} \gg V \  {\rm (interaction\ scale)} \gg W  \ {\rm (band\ width)} ,\label{projectioncriteria}\end{equation}
where it is reasonable to expect strongly correlated phases, such as FCIs, confined to a given topological band.
However, as we shall discuss later, a large $F$ is only one of several criteria for a 'good' topological flat band: the existence of FCIs is also crucially dependent on the structure of the Bloch states $\ket{\psi_s(\mathbf k)}$ (e.g., through the Berry curvature), the topology/coordination of the underlying lattice as well as the range of the interaction. See Section \ref{good} for a more detailed discussion.

There has been a flurry of recent works engineering topological flat band models with various benefits. To mention a few, there are useful $|C|=1$ models on the checkerboard lattice,\cite{chernins2,chernins3} the honeycomb lattice,\cite{chernins3,bosons,nonab2} the kagome lattice,\cite{chernins1} the ruby lattice,\cite{ruby} the triangular lattice,\cite{ifw2,ifw1} and the Kapit-Mueller construction\cite{kapit} (exact Landau level structure with long-range hopping) which can be formulated on various lattices. Models with $|C|=2$ flat bands were found on the dice lattice,\cite{c2} and in a triangular lattice model,\cite{ChernTwo} while systematic generalizations to any Chern number, $|C|=N$, are given by the pyrochlore slab model (multilayer kagome lattice model),\cite{max} multi-orbital square lattice models\cite{dassarma} and optical flux lattices\cite{opticalflux,coopermoessner} (formulated as hopping models in reciprocal space). Given the experience gained from the construction of these and other models it is straight-forward to create further examples on essentially any lattice.\footnote{The precise lattice structure is however often important for the topological properties of natural short-range models. This fact is well illustrated for time-reversal invariant models in Ref.\cite{dario} and for the models with variable Chern number in Refs.\cite{max,dassarma}.}

\begin{figure}[bt]
\centerline{\psfig{file=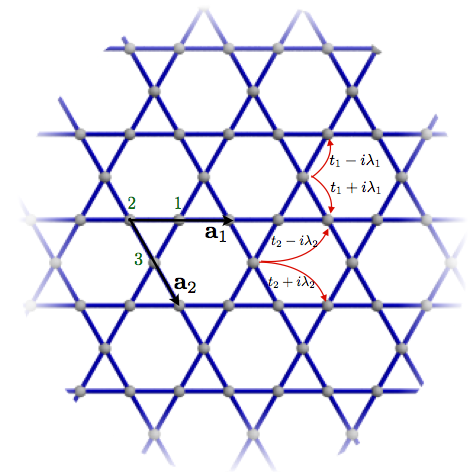,width=0.6\linewidth}}
\vspace*{8pt}
\caption{\textbf{The spin-orbit coupled kagome lattice model}. The unit cell has three sites (green numbers) and the Bravais lattice is generated by the lattice vectors $\mathbf a_1$  and $\mathbf a_2$. The red arrows indicate the considered nearest and next-nearest hopping processes, for which the sign of the spin-orbit terms, $\lambda_1,\lambda_2$, depend on the orientation of the process within a given hexagon (see the text). The Bloch Hamiltonian corresponding to this tight-binding model is given in Eq.~(\ref{blochham}).\label{kagome}}
\end{figure}

To illustrate the general concepts introduced above we will now go on to describe a three-band model with spin-orbit coupled particles on the kagome lattice\cite{chernins1} (Fig.~\ref{kagome}) in more detail.

\subsubsection*{Example: Kagome lattice model}

\begin{figure}[bt]
\centerline{\psfig{file=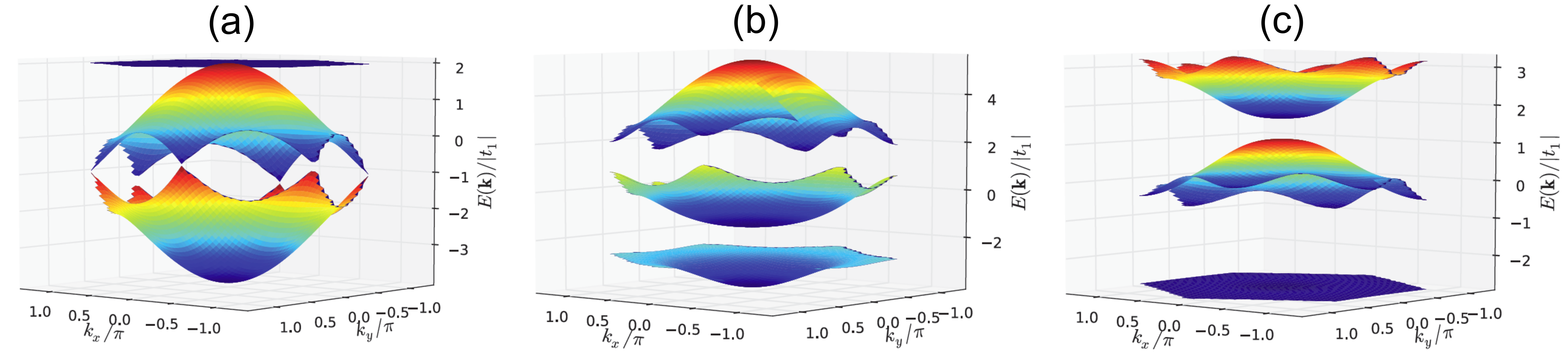,width=\linewidth}}
\vspace*{0pt}
\caption{\textbf{Band structure of the kagome model.} Energy spectra $E_s(\bb k); s=1,2,3$ for the kagome lattice model, defined in Eq.~(\ref{blochham}) for three different sets of tight-binding parameters: (a) with only real nearest neighbor hopping, $t_1=-1$, (b) also including complex nearest neighbor terms, $-t_1=\lambda_1=1$, (c) including also next-nearest neighbor terms, $t_1=-1, t_2=0.3,\lambda_1=0.28,\lambda_2=0.2$, leading to a flatness ratio of $F\approx52$ for the lowest band. }\label{kagomedispersion}
\end{figure}

To construct the Bloch Hamiltonian of the kagome model\footnote{The effective model only includes spin-polarized/spin-less particles. In electronic systems, this can be realized e.g. by proximity to ferromagnetic substrates or by applying a weak (or in-plane) Zeeman field.} , as illustrated in Fig.~\ref{kagome}, we note that the definition in Eq.~(\ref{bloch}) implies that hopping inside one unit cell does not give rise to a phase factor, while hopping a distance $\delta \bb R=n\bb a_1+m\bb a_2$ (to another unit cell) gives rise to a phase factor $e^{i\bb k\cdot \delta \bb R}$. With this convention, and $k_i=\bb k \cdot \bb a_i, \ i=1,2$, $k_3=k_1-k_2$, the Hamiltonian in the reciprocal space reads

\begin{align}
    \mathcal{H}_\bb k &=
    t_1
    \begin{pmatrix}
        0 & 1 + e^{i k_1} & 1 + e^{i k_2} \\
        1 + e^{- i k_1} & 0 & 1 + e^{-i k_3} \\
        1 + e^{- i k_2} & 1 + e^{i k_3} & 0 \\
    \end{pmatrix} \nonumber \\
    & + i \lambda_1
    \begin{pmatrix}
        0 & 1 + e^{i k_1} & -(1 + e^{i k_2}) \\
        -(1 + e^{- i k_1}) & 0 & 1 + e^{-i k_3} \\
        1 + e^{- i k_2} & -(1 + e^{i k_3}) & 0 \\
    \end{pmatrix}
    \;  \nonumber \\
    \; &+  t_2
    \begin{pmatrix}
        0 & e^{i k_2} + e^{i k_3} & e^{i k_1} + e^{-i k_3} \\
        e^{- i k_2} + e^{-i k_3} & 0 & e^{-i k_1} + e^{i k_2} \\
        e^{- i k_1} + e^{i k_3} & e^{i k_1} + e^{-i k_2} & 0 \\
    \end{pmatrix} \nonumber \\
    & + i \lambda_2
    \begin{pmatrix}
        0 & -(e^{i k_2} + e^{i k_3}) & e^{i k_1} + e^{-i k_3} \\
        e^{- i k_2} + e^{-i k_3} & 0 & -(e^{-i k_1} + e^{i k_2}) \\
        -(e^{- i k_1} + e^{i k_3}) & e^{i k_1} + e^{-i k_2} & 0 \\
    \end{pmatrix}\ .
    \label{blochham}
\end{align}
For $t_1<0, t_2=\lambda_1=\lambda_2=0,$ the spectrum of (\ref{blochham}) is built up by the well-known spectra of (spinless) graphene, including two Dirac cones, and, in addition, a perfectly flat band, which can neatly be understood in terms of localized modes\footnote{To see this, note that a state prepared with amplitudes of equal magnitude but alternating signs around a hexagon cannot leak out of the hexagon (all such amplitudes cancel) and the prepared state is therefore an eigenstate of the nearest neighbor hopping model.}, as shown in Fig.~\ref{kagomedispersion}(a). The flat band is not isolated---one of the dispersive bands has a quadratic dispersion around the touching point located at ${\bb k}=0$. The ambiguity of assigning a band index within the two-dimensional degenerate manifold of states at  ${\bb k}=0$ makes the Berry curvature ill-defined at this point and it is hence not meaningful to assign a Chern number to either of the bands.

Including finite spin-orbit coupling immediately opens a gap at the touching points and assigns a well defined Chern number for each of the bands---by adiabatic continuity these Chern numbers cannot be altered without a closing of the gap. In Figs.~\ref{kagomedispersion}(b,c) the lowest bands have $C=1$, the middle bands have $C=0$ and the upper bands have $C=-1$. For the band structure in Fig.~\ref{kagomedispersion}(b) we have used nearest neighbor hopping only which limits the flatness ratio $F\leq 1$ (we have $-t_1=\lambda_1=1$ which saturates the flatness limitation, i.e. $F=1$). In Fig.~\ref{kagomedispersion}(c), a very large flatness ratio, $F\approx52$, is obtained by including also next-nearest hopping ($t_1=-1, t_2=0.3,\lambda_1=0.28,\lambda_2=0.2$) which shows how rapidly the bands in this model can be flattened by including longer range terms (and fine-tuning).\cite{chernins1}

\begin{figure}[bt]
\centerline{\psfig{file=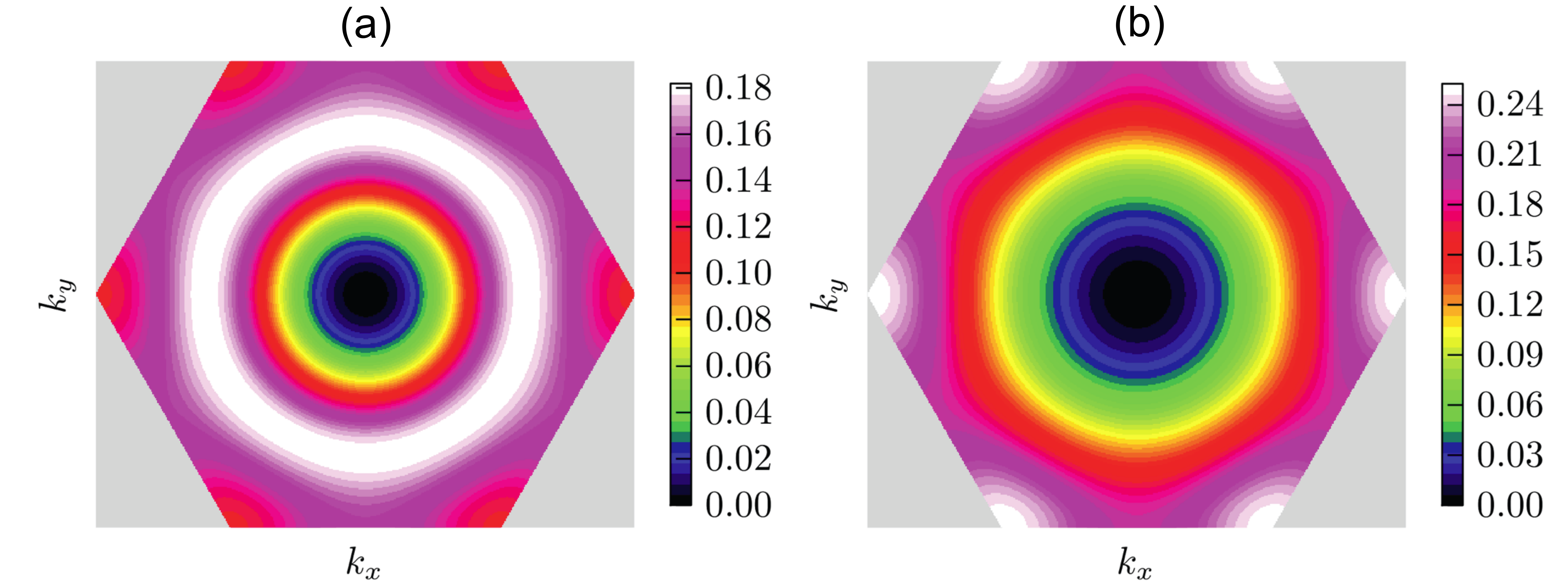,width=\linewidth}}
\vspace*{0pt}
\caption{\textbf{Berry curvature in the lowest kagome band} with (a) nearest neighbor hopping only ($-t_1=\lambda_1=1$) and (b) for the very flat band obtained by $t_1=-1, t_2=0.3,\lambda_1=0.28,\lambda_2=0.2$. These parameter sets are the same as for the dispersion shown in Figs.~\ref{kagomedispersion}(b,c). In both cases the integrated Berry curvature gives $C=1$, and one may note that the non-flat band corresponding to (a) actually has a slightly less peaked Berry curvature compared to the very flat band corresponding to (b) (standard deviation $\sigma_{F_{12}}=0.045$  vs. $\sigma_{F_{12}}=0.071$). In fact, numerical simulations indicate that the Berry curvature in (a) is more favorable for finding FCIs (assuming band flattening, cf. Section \ref{bandproj}).}\label{kagomeberry}
\end{figure}

In Figs.~\ref{kagomeberry}(a,b) we display the Berry curvature of the lowest band, which is readily computed numerically using Eq.~(\ref{berrycurvature2}), for the same parameter sets as used in Figs.~\ref{kagomedispersion}(b,c) respectively. These plots illustrate the general property that the Berry curvature in lattice models is necessarily inhomogeneous, which is especially prominent in models with a low number of bands (or equivalently, a low number of orbitals per unit cell).

In Section \ref{flatbandCN} we will use above results and show how the kagome model can be used as a building block for flat band models with arbitrary Chern numbers, by considering stacked kagome layers in the form of a pyrochlore slab.

\section{Interactions and band projection} \label{bandproj}

What makes the flat bands so interesting is that they amplify the effect of interactions. This Section provides detailed steps needed for describing interactions efficiently in the language of second quantization. To this end we consider the most natural (diagonal) lattice interactions of the form
\begin{equation}
H_\textrm{int}=\sum_{n,m,a,b}V^{ab}_{nm}c^\dagger_{n,a}c^\dagger_{m,b}c_{m,b}c_{n,a} \ .
\end{equation}
In reciprocal space this amounts to
\begin{equation}
H_\textrm{int}=\sum_{\substack{\bb k_1,\bb k_2,\bb k_3,\bb k_4\\ a,b}}V_{\bb k_1\bb k_2\bb k_3\bb k_4}^{ab}c^\dagger_{{\bb k_1},a}c^\dagger_{{\bb k_2},b}c_{{\bb k_3},b}c_{{\bb k_4},a} \ ,
\end{equation}
where the interaction matrix elements read
\begin{eqnarray}
V_{\bb k_1\bb k_2\bb k_3\bb k_4}^{ab}&=&\frac {\delta_{\bb k_1+\bb k_2,\bb k_3+\bb k_4}'} {N_s^2} \sum_{n,m} V^{ab}_{nm}e^{-i(\bb k_1-\bb k_4)\cdot (\bb R_n-\bb R_m)} \nonumber\\
&=&\frac {\delta_{\bb k_1+\bb k_2,\bb k_3+\bb k_4}'} {N_s} \sum_{n} V^{ab}_{n1}e^{-i(\bb k_1-\bb k_4)\cdot (\bb R_n-\bb R_1)},
\label{intm}
\end{eqnarray}
where $\delta'_{\bb k,\bb k'}$ is the two-dimensional periodic Kronecker delta function (with period $2\pi$) and we have again assumed translation invariance.
A number of numerical studies have managed to extract useful information about the full model
\begin{equation}
H=H_0+H_\textrm{int}=\sum_{a,b,{\bb k}} \mathcal H_\bb k^{ab}c^\dagger_{{\bb k},a}c_{{\bb k},b}+\sum_{\substack{\bb k_1,\bb k_2,\bb k_3,\bb k_4\\ a,b}}V_{\bb k_1\bb k_2\bb k_3\bb k_4}^{ab}c^\dagger_{{\bb k_1},a}c^\dagger_{{\bb k_2},b}c_{{\bb k_3},b}c_{{\bb k_4},a}  \ ,\label{unprojected}
\end{equation}
at various filling fractions including strong evidence for FCI phases, see e.g., Refs.\cite{chernins3,cherninsnum1,dassarma,ChernTwo,ifw2,bosons,nonab2,ifw1,disorder,ifwlong}. It should be noted that a corresponding option to numerically study the full problem is not available in the continuum FQH case---a discrete finite size problem only results after Landau level projection.

Nevertheless, to gain more insights about the underlying lattice physics, its relation to the continuum problem, and to motivate approximation schemes, it is useful to go on and introduce new (projected) creation operators, $d^\dagger_{{\bb k},s}$, such that
\begin{equation}
d^\dagger_{{\bb k},s}=
\sum_a \psi^a_s(\bb k) c^\dagger_{{\bb k},a} \ ; \ \  c^\dagger_{{\bb k},a}=\sum_{s} \psi^{a*}_s(\bb k) d^\dagger_{{\bb k},s} \ .\label{bandcreation}
\end{equation}
Clearly, $d^\dagger_{{\bb k},s}$ creates a particle with momentum ${\bb k}$ living in band $s$. In terms of these new operators one finds
\begin{equation}
H=\sum_{s,{\bb k}} E_s(\bb k)d^\dagger_{{\bb k},s}d_{{\bb k},s}+\sum_{\substack{\bb k_1,\bb k_2,\bb k_3,\bb k_4\\ s,s',s'',s'''}}\tilde V_{\bb k_1\bb k_2\bb k_3\bb k_4}^{ss's''s'''}d^\dagger_{{\bb k_1},s}d^\dagger_{{\bb k_2},s'}d_{{\bb k_3},s''}d_{{\bb k_4},s'''}, \ \label{unprojectedbandbasis}
\end{equation}
where
\begin{equation}
\tilde V_{\bb k_1\bb k_2\bb k_3\bb k_4}^{ss's''s'''}=\sum_{a,b}V_{\bb k_1\bb k_2\bb k_3\bb k_4}^{ab}\psi^{a*}_s(\bb k_1) \psi^{b*}_{s'}(\bb k_2) \psi^b_{s''}(\bb k_3) \psi^a_{s'''}(\bb k_4) \ .
\end{equation}
Eq.~(\ref{unprojectedbandbasis}) makes it evident that, in addition to the band preserving single particle terms, the interactions give rise to scattering between different bands. However, in the case of large band gaps compared to the interaction strength [cf. Eq.~(\ref{projectioncriteria})], it makes sense to project the problem onto the lowest partially filled band---this is where all the action takes place and it is the situation that we are primarily interested in. Furthermore, if the interaction is nevertheless much stronger than the bandwidth, an effective model of the form
\begin{equation}
H_{{\rm flat}}^{{\rm proj}, s}=\sum_{\bb k_1,\bb k_2,\bb k_3,\bb k_4} V^{{\rm proj}, s}_{\bb k_1\bb k_2\bb k_3\bb k_4}d^\dagger_{{\bb k_1},s}d^\dagger_{{\bb k_2},s}d_{{\bb k_3},s}d_{{\bb k_4},s}\ , \label{flatbandmodel}
\end{equation}
 with
 \begin{eqnarray}\!\!\!\!\!\!\!\!\!\! V^{{\rm proj}, s}_{\bb k_1\bb k_2\bb k_3\bb k_4}\!\!&\equiv & \tilde V_{\bb k_1\bb k_2\bb k_3\bb k_4}^{ssss}=\sum_{a,b}V_{\bb k_1\bb k_2\bb k_3\bb k_4}^{ab}\psi^{a*}_s(\bb k_1) \psi^{b*}_{s}(\bb k_2) \psi^b_{s}(\bb k_3) \psi^a_{s}(\bb k_4)\nonumber\\
 &=&\!\frac {\delta_{\bb k_1+\bb k_2,\bb k_3+\bb k_4}'} {N_s}\! \sum_{a,b,n} \! V^{ab}_{n1}e^{-i(\bb k_1-\!\bb k_4\!)\cdot (\bb R_n\!-\bb R_1\!)} \psi^{a*}_s(\bb k_1) \psi^{b*}_{s}(\bb k_2) \psi^b_{s}(\bb k_3) \psi^a_{s}(\bb k_4) ,\label{mael}\end{eqnarray}
describing interactions in the flattened band $s$, is well motivated.

There are several benefits of studying Eq.~(\ref{flatbandmodel}) compared to the full problem defined by Eq.~(\ref{unprojected}). In particular, it nicely isolates the effect of interaction, makes the study of considerably larger systems sizes tractable in numerics, and makes the comparison to the continuum FQH more explicit. After the initial work of Ref.\cite{cherninsnum2} (see also Ref.\cite{qi}), there have indeed been a number of successful works using the projected interactions leading to a deepened understanding of the connection between conventional Landau level physics and the interacting phase diagram in a partially filled flat Chern band, see e.g., Ref.\cite{andreas}. The projection has also been crucial in studies\cite{ChernN,ChernN2,BlochFCI} of interactions in bands with $|C|>1$.

We would also like to briefly mention that there is interesting work\cite{bands1,bands2,nonab1,roy,milica} on the algebraic structure obeyed by the projected density operators
\begin{equation}
\rho_{\bb q,s}=\sum_\bb k \Big[\sum_a e^{-i\bb q \cdot \delta\bb r_a}\psi_s^{a*}(\bb k) \psi_s^a(\bb k+\bb q)\Big]d^\dagger_{{\bb k},s}d_{{\bb k+\bb q},s}\ .
\end{equation}
In the long wave-length limit (small $|\bb q|$) one can make direct contact with the Girvin-MacDonald-Platzman (aka $W_\infty$ or GMP) algebra,\cite{gmp}
\begin{equation}[\rho_{\bb q_1,s},\rho_{\bb q_2,s} ]=2i\sin\Big(\frac{\bb q_1\times \bb q_2}{2} \ell^2\Big)\rho_{\bb q_1+\bb q_2,s}\ ,\label{gmpcom}\end{equation}
known to hold for the projected density operators in a Landau level.
For the Chern bands Eq.~(\ref{gmpcom}) holds in the limit of flat Berry curvature (and small $|\bb q_1|,|\bb q_2|$) which is increasingly realistic in models with larger unit cells. Here one is lead to the useful identification between the magnetic length, $\ell$, and the average Berry curvature, $\av{F^s_{12}({\mathbf k})}\propto C\propto \ell^2$ which highlights the fact that the Berry curvature plays a role similar to the magnetic field in the Landau level case. Further insights into the band geometry of Chern bands have been obtained by Roy\cite{roy} who stressed the importance of the Fubini-Study metric, which occurs in higher-order expansions (beyond linear in $|\bb q|$) of the density commutators [cf. Eq.~(\ref{gmpcom})]. It is worth notice that these seemingly quite abstract quantities have direct experimental consequences---while the connection between the Berry curvature and the (anomalous) Hall effect has been discussed above, it has also been suggested that the (symmetric part of the) Fubini-Study metric tensor also has measurable consequences, e.g., in the current noise spectrum.\cite{fubininoise} For a more in depth discussion of the algebraic properties of Chern bands we refer to the recent review in Ref.\cite{otherreview} and the references therein.

\section{Interactions in $|C|=1$ models and the FQH analogy}\label{c1}

In this Section we consider the effect of interactions within flat bands with unit Chern number. In particular, we focus on the analogy with conventional Landau level physics as well as new competing phases that are specific to the lattice setting.

\subsection{Basic identification of FCIs}
A standard approach to uncover interaction-induced new insulating phases in flat bands is to extract the low-energy physics of (\ref{flatbandmodel}) on a finite system---$N_e$ electrons (or $N_b$ bosons) in a lattice with $N_1$ and $N_2$ unit cells in two primary directions ($N_s=N_1\times N_2$), by numerical algorithms such as exact diagonalization and density matrix renormalization group (DMRG). A pressing question immediately arises: how can we actually identify FCI phases based on numerical data?

Generally speaking, the most obvious numerical evidence of FCIs is the topological degeneracy of ground states on the torus. At band filling $\nu=N_e/N_s=p/q$, where $p$ and $q$ are coprime, at least $q$ (quasi-) degenerate ground states are expected\footnote{The number of quasi-degenerate states is equal to $q$ for Abelian FCIs, but larger than $q$ for non-Abelian FCIs. In Landau levels all eigenstates of a translation invariant operator are at least $q$-fold degenerate\cite{Haldane85}. In Chern bands, however, a $q-$fold quasi-degeneracy is nontrivial.}. The ground-state manifold should be separated from excited levels by a many-body gap that does not vanish in the thermodynamic limit and, ideally, the ground state splitting should vanish exponentially with system size. Moreover, the topological degeneracy should be robust to the twisted boundary conditions, which can be demonstrated by calculating the spectral flow (i.e. the energy spectra as a function of boundary conditions): the ground states should never mix with the excited levels under the insertion of magnetic flux $\Phi$ through the handle of the torus. Typically, the ground states evolve into each other in the spectra flow for an appropriate flux insertion, which also suggests a non-trivial quantized Hall conductance.

Because of the translation invariance in the lattice, the total momentum $\mathbf K=(K_1,K_2)$ is a good quantum number [this can also be seen from the Hamiltonian matrix elements (\ref{mael})] and provides convenient labels for the many-body states. In most cases (but not always), one can simply deduce the momentum sectors where the FCIs are expected to occur by first folding $\mathbf K$ into $K_{1\textrm{D}}=K_1+N_1 K_2$\cite{cherninsnum2} and then applying an exclusion rule known from the thin-torus limit\cite{bklong} (or equivallently, "root configuration"\cite{rootc} or "pattern of zeros"\cite{pattern}) of the corresponding FQH states, which implies that there are no more than $p$ particles in $q$ consecutive orbitals at, and slightly below, $\nu=p/q$. A technically more involved prescription was proposed in Ref.\cite{nonab1}, by which one can precisely predict the ground-state momenta as well as the quasihole counting in each momentum sector of the FCIs, which can be compared with the numerical data to examine whether the numerical ground states are indeed in the FCI phase.

Besides the energetic evidence, entanglement measures, especially the particle-cut entanglement spectrum\cite{pes} can provide more insights into the identification of the ground states.\cite{cherninsnum2} The quasihole excitation properties reflected by the particle-cut entanglement spectrum can be used as an indication of the topological order and sometimes to rule out other competing phases. A complementary way to study the quasihole excitations is, quite naturally, to change the lattice size so that $N_e<\nu N_s$, which can add holes into the system.

\begin{figure}[bt]
\centerline{\psfig{file=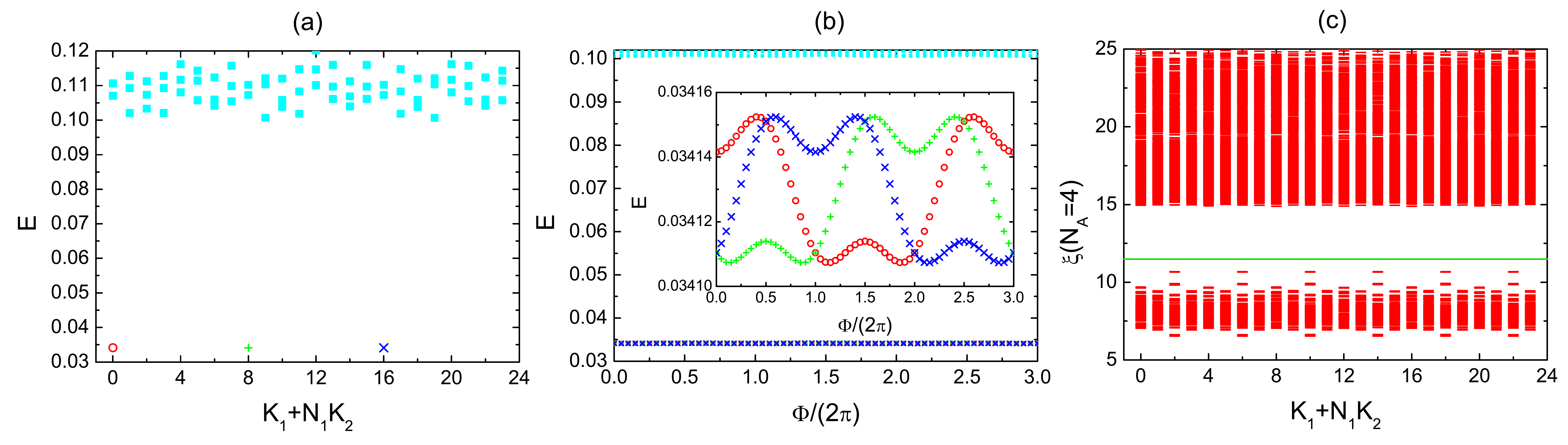,width=5in}}
\vspace*{0pt}
\caption{\textbf{Numerical observation of a FCI in a $C=1$ band.} $N_e=8$ interacting electrons projected to the flattened $C=1$ band ($-t_1=\lambda_1=1$) on a kagome lattice with $N_1\times N_2=4\times 6$ unit cells. Note that the underlying kagome lattice has $N_1\times N_2\times N_c=4\times 6\times 3=72$ sites and the band filling is $\nu=N_e/(N_1\times N_2)=1/3$. (a) shows the low lying energies with periodic boundary conditions (no flux) in each (center of mass) momentum sector. There is a three-dimensional manifold of quasi-degenerate states (colored red, green, blue for clarity) whose momenta can be deduced from a simple (one particle in three consecutive orbitals) exclusion rule
 in close analogy with the FQH exclusion rules. (b) shows the spectral flow under flux insertion (see the main text). On the scale of the main plot all three ground states appear to be degenerate throughout, but zooming as shown in the inset reveals that the ground states flow into each other. This spectral flow, where the ground states evolve without mixing with the higher states also makes a strong case for the quantized conductance, $\sigma_{xy}=\nu\frac{h}{e^2}$. (c) The ground-state particle-cut entanglement spectrum with a clear entanglement gap. The total number of levels below the gap is the same as the corresponding quasihole excitation counting of the $\nu=1/3$ FQH Laughlin state.}
\label{kagome8electrons}
\end{figure}

As a typical example, the numerical data for $N_e=8$ interacting electrons at $\nu=1/3$ with a nearest neighbor repulsion projected to a flattened band in the $N_1\times N_2=4\times 6$ kagome lattice model is shown in Fig.~\ref{kagome8electrons}. The excellent finite size three-fold ground-state quasi-degeneracy and the large gap are characteristic for the $\nu=1/3$ FCI which is known to generally be the most stable electronic state, and moreover, the kagome lattice model stands out as a good lattice models for realizing several $C=1$ FCIs, see e.g., the discussion by Wu {\it et~al.}\cite{nonab3} and Sections \ref{numobs}-\ref{good}. The three quasi-degenerate ground states evolve into each other without mixing with exited levels during the spectral flow, which clearly shows the robustness of topological degeneracy and strongly suggests the quantized Hall conductance $\sigma_{xy}=\frac{1}{3}\frac{h}{e^2}$. The total number of levels below the clear entanglement gap in the particle-cut entanglement spectrum matches the counting of $\nu=1/3$ FQH Laughlin quasihole excitations, indicating the ground states are indeed topological nontrivial and belong to the same phase as the $\nu=1/3$ FQH Laughlin state.

Further evidence used to confirm that the ground states are FCIs, include the adiabatic continuity to the FQH states (as discussed later in Section \ref{details}) and the modular matrices that contain the information of statistics of quasiparticles.\cite{modularfci,modularfci2}

\subsection{Numerically observed states}\label{numobs}

Based on numerical evidence described above, a large number of FCIs have been observed in $|C|=1$ models\footnote{The numerical criteria described above to identify FCIs also work in $|C|>1$ models.}. They appear as the lattice analogues of the most well-known FQH states, such as the Laughlin,\cite{Laughlin83} composite fermion (CF),\cite{jain} Moore-Read (MR),\cite{mr} and Read-Rezayi (RR)\cite{rr} states.

The Laughlin analogues at $\nu=1/m$ are the first FCIs that were discovered. In the initial papers of FCIs, $\nu=1/3$ and $\nu=1/5$ fermionic states in the checkerboard lattice model were reported,\cite{chernins3,cherninsnum1,cherninsnum2} and subsequent works\cite{nonab3} confirmed the $\nu=1/3$ fermionic states in various lattice models, such as the honeycomb lattice model, two-orbital square lattice model, kagome lattice model and ruby lattice model. Besides an early work in the Kapit-Mueller model,\cite{kapit} the $\nu=1/2$ bosonic states were also observed in the honeycomb lattice model and checkerboard lattice model for two-body hard-core bosons.\cite{bosons}

After the discovery of FCIs at Laughlin filling fractions, $\nu=1/q$, the members of the FCI family naturally extend to CF states at $\nu=n/(2n+1)$ for fermions and $\nu=n/(n+1)$ for bosons. The existence of such states was briefly mentioned in Ref.\cite{ifw2}, and convincing evidence of CF FCIs for fermions in the checkerboard lattice model was provided in Ref.\cite{andreas} while Ref.\cite{CompositeFCI} reported compelling indications of bosonic CF FCIs in the ruby lattice model.

Beyond Abelian states, non-Abelian FCIs at $\nu=k/(k+2)$ for fermions and $\nu=k/2$ for bosons $(k>1)$ were also observed. Initially, $(k+1)$--body interactions seemed necessary to stabilize these states, such as $\nu=1/2$ fermionic MR FCI and $\nu=3/5$ fermionic RR FCI in the checkerboard lattice model,\cite{nonab1} $\nu=1/2$ fermionic MR FCI in the kagome lattice model,\cite{nonab3} and $\nu=1$ bosonic MR FCI in the honeycomb lattice model for three-body hard-core bosons.\cite{nonab2} More recently, it has been shown that some non-Abelian FCIs can also appear with significantly more realistic two-body interactions. Tentative evidence was first provided for the $\nu=1$ bosonic MR FCI in the optical flux lattice model with a simple on-site repulsion.\cite{opticalflux} Stronger evidence, including finite size scaling of gaps, has been found is a so-far unpublished work for the $\nu=1$ bosonic MR FCI as well as for the $\nu=3/2$ bosonic RR FCI in the Kapit-Mueller model with longer range interactions.\cite{eliotzhao} Moreover, it has been shown that a confining potential can significantly help to stabilize non-Abelian FCIs.\cite{dmrg}

Most of the above FCIs are found for short-range interactions, i.e. between nearest neighbor sites for fermions and onsite in the case of bosons. Inclusion of longer range terms, such as Coulomb or dipolar interactions typically weaken, or destroy, the Abelian FCIs. However, in certain cases, non-Abelian FCIs can be stabilized by longer-range interactions.\cite{eliotzhao}

\subsection{Wannier mapping -- a bridge between FCIs and FQH states} \label{wannier}

The similarity of emergent features suggests that the observed FCIs in $|C|=1$ models and the corresponding FQH states are in the same topological phase. However, FCIs and FQH states exist in two markedly different systems. It is a natural question  how well the FCIs described by model wave functions which have historically been instrumental for the understanding of the FQH effect. The key procedure for answering this question is to find a basis of one-particle states in Chern bands and in Landau levels respectively, so that these two basis have similar properties and can be mapped to each other. Then, the FCI and FQH problem can be put on equal footing and quantitatively compared. In Ref.\cite{qi}, Qi constructed the Wannier states in Chern bands that can mimic the Landau gauge one-particle states in Landau levels, and this construction was generalized by Wu {\em et. al.} to a gauge-fixed version appropriate for numerical finite size studies.\cite{gaugefixing} As discussed in detail below, the mapping between the Wannier states and the Landau gauge one-particle states is greatly helpful for the direct comparison of FCI and FQH bulk wave functions, the establishment of adiabatic continuity, and even the investigation of the edge physics in FCI systems.

\subsubsection{Wannier states construction}
It is well known that it is not possible to construct a complete basis set of exponentially localized wave functions for a band with $C\neq 0$, see e.g Ref.\cite{LLlocal}. In fact, in any system supporting a Hall current, the orbitals cannot (asymptotically) decay faster than $1/r^2$ as shown by Thouless.\cite{hallabsence} Precisely this absence of a basis of localized states is at the heart of why FQH (and FCI) states can at all be favored over Wigner crystals although the kinetic energy is entirely quenched\footnote{In the QH systems Wigner crystals are nevertheless expected to be favored at very low filling fractions, $\nu\lesssim 1/7$.\cite{lam}}.

However, in Chern bands as in the case of Landau levels, it is generally possible to construct basis states that are localized in one of the two spatial directions while being delocalized in the other direction.\cite{qi,wannier_review,sv} To make the discussion as transparent as possible, we consider a system with a simple rectangular unit cell and refer the interested reader to existing literature for more general constructions.\cite{gaugefixing,AdiabaticContinuity1,AdiabaticContinuity3,wanniersym}
The states
\begin{eqnarray}
\psi_{p,j}^{\textrm{torus}}(x,y)&=&\sum_{m=-\infty}^{+\infty}\psi_{p,j+mN_{\phi}}^{\textrm{cylinder}}(x,y)
\label{psikLL}
\end{eqnarray}
with
\begin{eqnarray}
\psi_{p,j}^{\textrm{cylinder}}(x,y)&=&\left(\frac 1 {\sqrt{\pi}L_2}\right)^{\frac{1}{2}}H_p\Big(x-\frac{2\pi}{L_2}j\Big)
e^{i\frac{2\pi}{L_2}jy}
e^{-(x-\frac{2\pi}{L_2}j)^2/2}
\end{eqnarray}
form a basis of one-particle states localized in the $x-$direction (usually called as orbital basis) in the $p$th Landau level on the torus, where $L_2$ is the circumference in the $y-$direction, $N_{\phi}$ is the number of magnetic flux quanta penetrating the torus, and $H_p, p=0,1,2,\ldots$ are the Hermite polynomials.

\begin{figure}[bt]
\centerline{\psfig{file=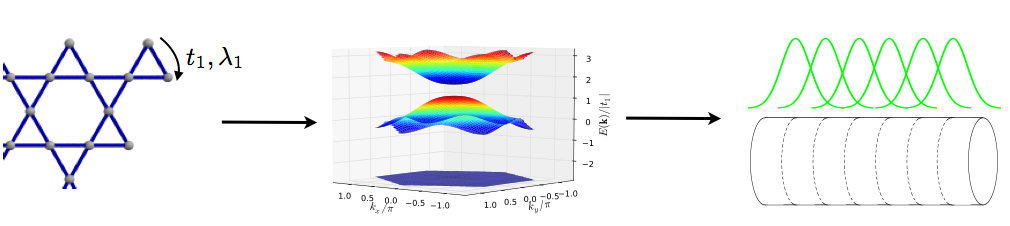,width=\linewidth}}
\vspace*{0pt}
\caption{\textbf{Wannier mapping from Chern bands to Landau levels.} The Wannier states, living in a Chern band of a tight-binding model, have similar localization properties as the orbitals in a Landau level, enabling a one-to-one mapping between the two systems.}\label{wannierconstruction}
\end{figure}

Following Qi,\cite{qi} analogous Wannier functions describing the Chern band in the continuum can be constructed on the torus by introducing
\begin{equation}
\ket{W_s(k_2,x)}= \frac 1 {\sqrt{L_1} }\sum_{k_1}e^{-i\int_0^{k_1} A^s_1(p_1,k_2)dp_1-ik_1[x-\theta(k_2)/2\pi]}\ket{\psi_s(\mathbf k)}\ ,\label{wannierwfs}
\end{equation}
where $L_1$ is the circumference in the $x-$direction, $\theta(k_2)=\int_0^{2\pi}A^s_1(p_1,k_2)dp_1$ and the gauge is chosen so that $A^s_2({\mathbf k})=0$. The states in Eq.~(\ref{wannierwfs}) are defined such that they are maximally localized in the $x-$direction similar to the Landau level orbitals (Fig.~\ref{wannierconstruction}).
It is useful to think of Eq.~(\ref{wannierwfs}) as a partial Fourier transform of
the momentum eigenstates which is additionally taking into
account the parallel transport resulting from the Berry connection.

The center-of-mass position of these states is
\begin{equation}
\av{\hat{x}}=\bra{W_s(k_2,x)}\hat{x}\ket{W_s(k_2,x)}=x-\theta(k_2)/2\pi\ ,
\end{equation}
which motivates the interpretation of $\theta(k_2)/2\pi$ as the "charge polarization".
Considering $C=-\frac 1 {2 \pi}[\theta(k_2+2\pi)-\theta(k_2)]$ which follows from Eq.~(\ref{chernnumber}) invoking the gauge condition $A^s_2({\mathbf k})=0$, we have $\av{\hat{x}}\rightarrow \av{\hat{x}}+C$ when $k_2\rightarrow k_2+2\pi$. Thus we can regard
the center of mass position of $\ket{W_s(k_2,x)}$ as a continuous function of $\widetilde{K}_2$, where $\widetilde{K}_2\equiv 2\pi x+Ck_2$ is a pseudomomentum with $k_2\in[0,2\pi)$. Noticing that the center of mass of Eq.~(\ref{psikLL}) also depends on a pseudomomentum $2\pi j/L_2$, we can tentatively build a mapping between $\ket{W_s(k_2,x)}$ and $|\psi_{p,j}^{\textrm{torus}}\rangle$: $\ket{W_s(k_2,x)}\leftrightarrow |\psi_{p,j}^{\textrm{torus}}\rangle$ with $\widetilde{K}_2=2\pi x+Ck_2=2\pi j/L_2$.
Then, the FCIs and FQH states can be compared by expanding them in the Wannier basis $\{\ket{W_s(k_2,x)}\}$ and orbital basis $\{|\psi_{p,j}^{\textrm{torus}}\rangle\}$, respectively.

At the conceptual level, the mapping established above is very satisfying. However, upon closer inspection, it suffers from two problems that are important to tackle before making quantitative comparisons between FCI and FQH states. It is important to notice that, as it stands, Eq. (\ref{wannierwfs}) is not uniquely defined due to a remaining gauge degree of freedom in choosing the Bloch states: $\ket{\psi_s(\mathbf k)}\rightarrow e^{i\phi(k_1)}\ket{\psi_s(\mathbf k)}$ can change $A^s_1({\mathbf k})$ without breaking the gauge condition $A^s_2({\mathbf k})=0$.
It is unclear which gauge is the most suitable one for implementing the mapping consistently\footnote{We have $\theta(k_2)\rightarrow\theta(k_2)+\phi(2\pi)-\phi(0)$ and
$\ket{W_s(k_2,x)}\rightarrow e^{i\phi(0)}\ket{W_s(k_2,x-[\phi(2\pi)-\phi(0)]/(2\pi))}$, under the gauge transform $\ket{\psi_s(\mathbf k)}\rightarrow e^{i\phi(k_1)}\ket{\psi_s(\mathbf k)}$.
}.
Moreover, the maximally localized Wannier states are not orthogonal in a finite system.
A direct application of Qi's mapping mapping therefore usually leads to small overlaps between the exact diagonalization ground states in the lattice and the FQH model states, underscoring the need for a solution of the gauge and orthogonality problems (see however Ref.\cite{AdiabaticContinuity1} for a successful application thereof). These issues were indeed considered in some detail in Ref.\cite{gaugefixing} where an efficient and systematic (albeit technically involved) procedure for obtaining well behaved gauge-fixed Wannier states that can be used for direct numerical comparison between FCIs in generic lattice models and FQH model states was obtained. In essence, the work of Ref.\cite{gaugefixing} traded the maximal localization of the Wannier states for orthogonality in finite systems, and provided an explicit recipe for this purpose.

Using the gauge fixed version of the Wannier mapping, we can successfully compare the FCIs as exact diagonalization ground states in lattice models with FQH model states, establish the adiabatic continuity between them, and study the FCIs by similar methods to those developed for FQH states (such as the orbital-cut entanglement spectrum\cite{LiH}).\cite{AdiabaticContinuity3}

A natural question is, given that Eq.~(\ref{wannierwfs}) as well as its gauge-fixed version in finite systems corresponds to the Landau-level-like wave-functions (\ref{psikLL}), then for which Landau level? The numerics discussed below clearly hints that they typically, at least for short-range interactions, correspond to the lowest Landau level, signaled e.g., by very impressive overlaps between the FCI wave functions in the gauge-fixed Wannier basis and the FQH model states in the lowest Landau level orbital basis.\cite{AdiabaticContinuity3,gaugefixing}

\subsubsection{Adiabatic continuity and bulk-edge correspondence}\label{details}
In the Wannier basis, two-body interactions projected to the flat band labeled by $s$ [Eq.~(\ref{flatbandmodel})] can be written as
\begin{eqnarray}
\widetilde{H}_{\textrm{flat}}^{\textrm{proj},s}=\sum_{j_1,j_2,j_3,j_4=0}^{N_s-1}
\widetilde{V}_{j_1j_2j_3j_4}^{\textrm{proj},s}d_{j_1,s}^\dagger d_{j_2,s}^\dagger d_{j_3,s} d_{j_4,s},
\label{fciham}
\end{eqnarray}
where $d_{j,s}^\dagger$ ($d_{j,s}$) creates (annihilates) a particle in the $j$th Wannier orbital of band $s$, and $\widetilde{V}_{j_1j_2j_3j_4}^{\textrm{proj},s}$ is nonzero only if $j_1+j_2=j_3+j_4$~(mod $N_2$).
This form is very similar to that of a FQH two-body Hamiltonian in the lowest Landau level:
\begin{eqnarray}
H_{\textrm{FQH}}=\sum_{j_1,j_2,j_3,j_4=0}^{N_\phi-1}
V_{j_1j_2j_3j_4}^{\textrm{FQH}}c_{j_1}^\dagger c_{j_2}^\dagger c_{j_3} c_{j_4},
\label{fqhham}
\end{eqnarray}
where $c_j^\dagger$ ($c_j$) creates (annihilates) a particle in the $j$th orbital of the lowest Landau level and $V_{j_1j_2j_3j_4}^{\textrm{FQH}}$ is nonzero only if $j_1+j_2=j_3+j_4$~(mod $N_\phi$).

Although possessing different symmetries, these two Hamiltonians have the same structure of the Hilbert space if $N_s=N_\phi$, so that an interpolating Hamiltonian
\begin{eqnarray}
H(\lambda)=\lambda w_{\textrm{FCI}}\widetilde{H}_{\textrm{flat}}^{\textrm{proj},s}+(1-\lambda)w_{\textrm{FQH}}H_{\textrm{FQH}}
\label{interpolate}
\end{eqnarray}
is well defined with $\lambda\in[0,1]$ the interpolation parameter.
Here $w_{\textrm{FCI}}$ and $w_{\textrm{FQH}}$ are energy rescaling factors
to make the energy gap at $\lambda=0$ and $\lambda=1$ equal to 1\footnote{In the cases where the adiabatic continuity holds it is well established that the gap survives in the thermodynamic limit at $\lambda=0$ and $\lambda=1$ respectively.}.

\begin{figure}[bt]
\centerline{\psfig{file=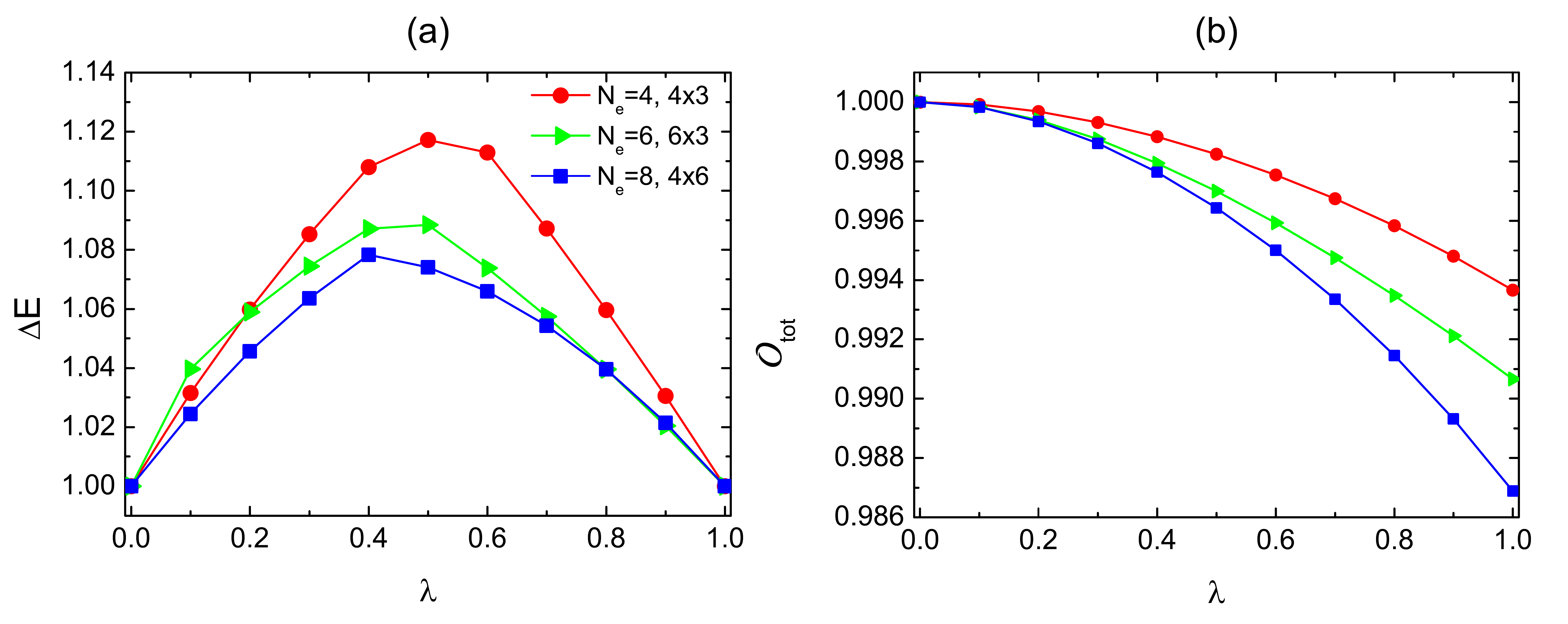,width=\linewidth}}
\vspace*{0pt}
\caption{\textbf{Adiabatic continuity between FCIs and FQH states.} Results of the interpolation Eq.~(\ref{interpolate})
for electrons at $\nu=1/3$ with $N_e=4$ (red dot), $N_e=6$ (green triangle),
and $N_e=8$ (blue square). In the FCI part, the lattice size is $N_1\times N_2=4\times3$,
$N_1\times N_2=6\times3$ and $N_1\times N_2=4\times6$, respectively; and
$-t_1=\lambda_1=1$.
(a) The energy gap $\Delta E$ does not close for any intermediate $\lambda$.
(b) The total overlap $\mathcal{O}_{\textrm{tot}}$ is still close to 1 at $\lambda=1$.
These results indicate that the adiabatic continuity holds for fermions at $\nu=1/3$.
}
\label{continuity}
\end{figure}

The adiabatic continuity between FCIs and FQH states requires that for any intermediate $\lambda$, the energy gap of $H(\lambda)$ remains finite in the thermodynamic limit. This requirement is satisfied for fermions at $\nu=1/3$ in the kagome lattice model, hence firmly establishing the adiabatic continuity between the FCI phase (the exact diagonalization ground states in the lattice, as shown in Fig.~\ref{kagome8electrons}) and FQH Laughlin model states (Fig.~\ref{continuity}). When $\lambda$ varies from 0 to 1, there are always three nearly-degenerate states separated from excited levels by a sizable energy gap $\Delta E$ which does not vanish. The total overlap $\mathcal{O}_{\textrm{tot}}=\frac{1}{3}\sum_{i=1}^3 |\langle\Psi^i(\lambda)|\Psi^i_{\textrm{FQH}}\rangle|^2$ between the (quasi-) degenerate ground states $|\Psi^i(\lambda)\rangle$ of $H(\lambda)$ and FQH Laughlin model states $|\Psi^i_{\textrm{FQH}}\rangle$ is close to 1 for any intermediate $\lambda$, which further corroborates that the ground states do not undergo a phase transition during the interpolation. The same analysis can be done at other filling fractions so that the adiabatic continuity between FCIs and FQH states can also be established for $\nu=1/2$ fermionic Moore-Read phase, $\nu=1/2$ bosonic Laughlin phase, and $\nu=1$ bosonic Moore-Read phase.\cite{AdiabaticContinuity3} In a somewhat different vein, the adiabatic continuity between FCI and FQH phases has also been argued to hold by passing to the Hofstadter model and continuing it to the low flux limit.\cite{AdiabaticContinuity2}

\begin{figure}[bt]
\centerline{\psfig{file=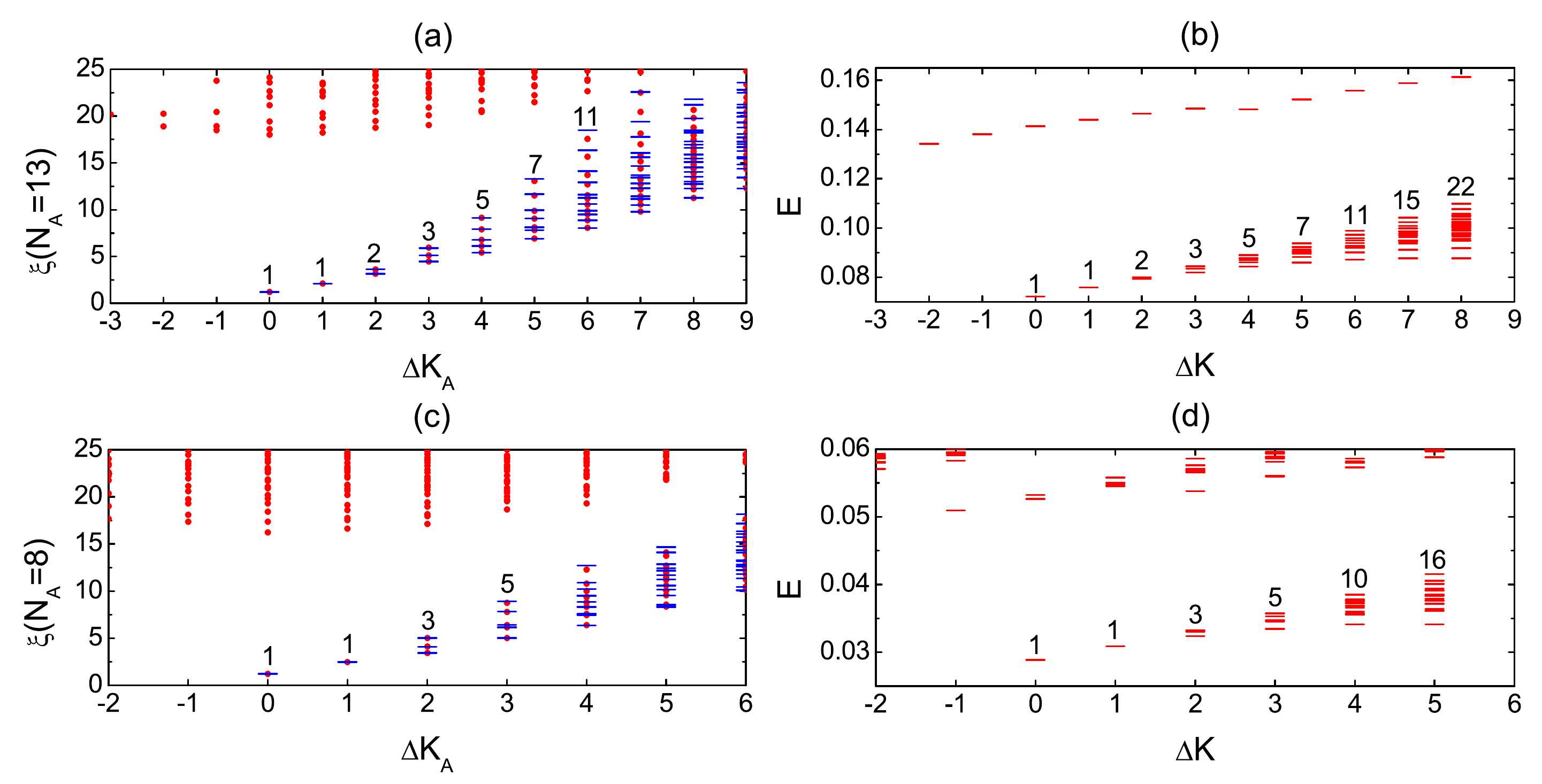,width=1.0\linewidth}}
\vspace*{0pt}
\caption{\textbf{Bulk-edge correspondence.} (a) The orbital-cut entanglement spectra of the bosonic FQH Laughlin model state (blue dashes) and the Laughlin FCI (red dots) at $\nu=1/2$ for 25 bosons. The size of the Ruby lattice is $N_1\times N_2=7\times7$.
(b) The edge excitation spectrum of the FCI system at $\nu=1/2$ for 8 bosons. The size of the Ruby lattice is $N_1\times N_2=5\times5$.
(c) The orbital-cut entanglement spectra of the bosonic FQH Moore-Read model state (blue dashes) and the Moore-Read FCI (red dots)
at $\nu=1$ for 16 bosons. The size of the Ruby lattice is $N_1\times N_2=3\times5$.
(d) The edge excitation spectrum of the FCI system at $\nu=1$ for 10 bosons. The size of the Ruby lattice is $N_1\times N_2=5\times3$.
}
\label{cylinder_OES}
\end{figure}

In addition to the establishing the adiabatic continuity between bulk states, the Wannier mapping is also useful to study the bulk-edge correspondence in FCI systems.\cite{dmrg} By appropriately generalizing the Wannier interaction matrix elements in Eq.~(\ref{fciham}) to a finite cylinder setup\footnote{In the Wannier setup, the cylinder matrix elements are exponentially localized with a localization length proportional to the circumference of the cylinder and they coincide with the corresponding torus matrix elements in the limit of a long torus.}, it is possible to utilize the advantages\footnote{A bipartitioning on the cylinder involves a single spatial cut rather than two as is natural on the torus. This enables the study of a single edge in the orbital-cut entanglement spectrum, and it is very favorable for application of entanglement based algorithms such as DMRG due to the lower ($\approx$ half) entanglement resulting from the single cut.\cite{dmrg}} of the cylinder geometry to investigate the correspondence between the bulk entanglement and edge excitations of FCIs in analogy to FQH systems. In Ref.\cite{dmrg} it was demonstrated that the counting structure in the orbital-cut entanglement spectrum is the same as in the edge excitation spectrum, thereby establishing the nontrivial bulk-edge correspondence of both Abelian [Figs.~\ref{cylinder_OES}(a,b)] and non-Abelian FCIs [Figs.~\ref{cylinder_OES}(c,d)].

A so-far less explored alternative to the Wannier state mapping is to instead transfer the FQH problem to the Bloch basis. In Ref.\cite{BlochFCI}, the authors first constructed a Bloch-like lowest Landau level basis by appropriately superposing the usual orbital basis states in Eq.~(\ref{psikLL}). Then, this new basis is mapped to the Bloch basis in the $C=1$ Chern band. This 'inverse' mapping from FQH problem to the Bloch basis provides an alternative way to compare the FCIs with FQH states, making use of more symmetries for numerical simulations compared to the gauge-fixed Wannier mapping.

\subsection{Pseudopotential analogy}\label{pseudop}
As discussed above, the existence of various FCIs in $|C|=1$ models was well established already in early numerical works. However, a systematic explanation for why just some certain states appear for a specific interaction would still be highly rewarding. An important step in that direction was taken by L\"auchli {\it et~al.},\cite{andreas} who developed a heuristic based on the analogy with Haldanes pseudopotentials,\cite{haldane83} which is a well-esthablished tool in the continuum Landau levels\footnote{Ref.\cite{wannierpp} attempted to extract the pseudopotential parameters (semi-)analytically using the Wannier basis. Unfortunately, however, it seems that the outcome thereof lacks predictive power in its present formulation.}.

To make the discussion self-contained, we start by considering a single Landau level and write the Hamiltonian describing any two-body interaction projected to a Landau level in terms of Haldane's pseudopotentials:\cite{haldane83}
 \begin{eqnarray}H=\sum_{i<j}\sum_{m=0}^{\infty}\mathcal V_mP_m^{ij},\label{pseudohamiltonian}\end{eqnarray}
where $P_m^{ij}$ projects onto a state where particles $i,j$ have
relative angular momentum $m$ and $\mathcal V_m$ is the pseudopotential
parameter, which is a real number determined by the specific
interaction. More generally, the pseudopotentials project onto components of the many-body wave function with certain vanishing properties and can as such be defined in generic (non-rotationally invariant) geometries such as the torus,\cite{haldanewfs} where the (yet unprojected) interaction takes the form
\begin{eqnarray}H=\sum_{i<j}\sum_{m=0}^{\infty}\mathcal V_mL_m(-\nabla_{i}^2)\delta(\mathbf r_i-\mathbf r_j).\nonumber\end{eqnarray} For any realistic interaction projected to  $n$th Landau level one can obtain the pseudopotential parameters as  \begin{equation}\mathcal V_m=\int
\mathcal F_n(q)V(q)L_m(q^2)e^{-q^2}\mathrm{d}^2{\mathbf q},\label{Vm}\end{equation} where
$V(q)$ is the Fourier transform of the
interaction potential
and $L_m$ is the Laguerre
polynomial with $L_0(q^2)=1, L_1(q^2)=1-q^2, \ldots$ and the Landau level form factor is $\mathcal F_n(q)=[L_n(q^2/2)]^2$.\footnote{In the relativistic case of graphene this is modified to $\mathcal F_n(q)=[L_n(q^2/2)+L_{n-1}(q^2/2)]^2/4$.\cite{graphenepp}} Once the dominant pseudopotential parameters are known, it is often possible to refer to the literature for candidate ground states. In fact, the most prominent FQH states---the Laughlin states at $\nu=1/q$---are exact and unique highest density zero energy eigenstates for $\mathcal V_m>0, m\leq q-2$ and  $\mathcal V_m=0, m> q-2$.\footnote{In the non-Abelian cases this scenario is generalized to multi-body pseudopotentials.\cite{pseudomulti}}

Alternatively, one can also numerically extract the pseudopotential parameters, $\mathcal V_m$, directly from the energy spectrum of two-interacting particles without using the analytical formula Eq.~(\ref{Vm}). In a Landau level on the torus, there are $2N_\phi$ finite energy levels for each pseudopotential and their values coincide with the respective pseudopotential parameters\footnote{Finite-size corrections (level splitting) occur if the torus is too thin (in either direction). The projector property of the pseudopotentials is however insensitive to this.}. This immediately generalizes to the Chern band and provides a way to extract analogues of the pseudopotential parameters for lattice systems despite the lack of an analytical expression like Eq.~(\ref{Vm}).\cite{andreas}

\begin{figure}[bt]
\centerline{\psfig{file=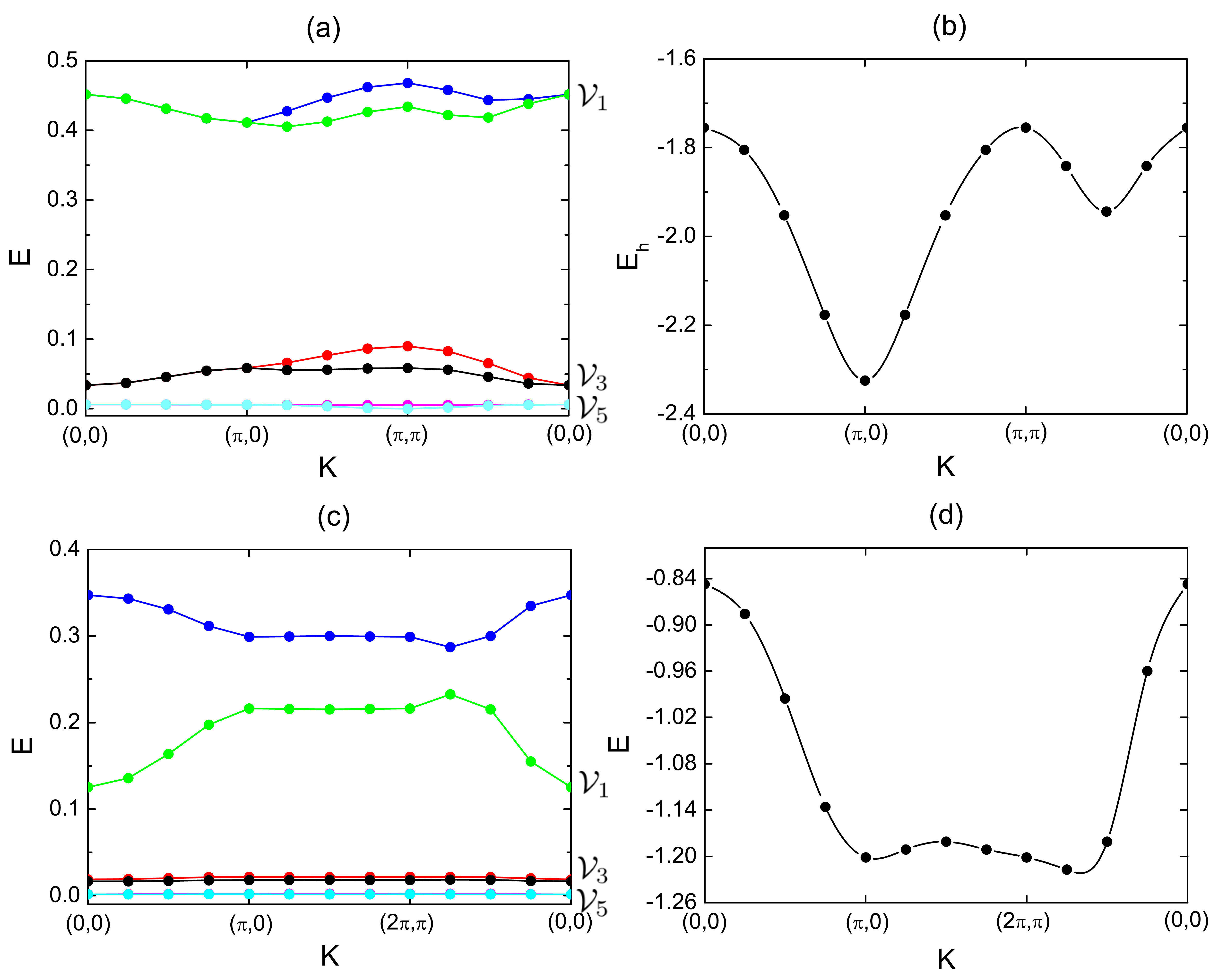,height=8cm,width=1.0\linewidth}}
\vspace*{0pt}
\caption{\textbf{Two-particle spectrum and single-hole dispersion} in (a,b) checkerboard lattice and (c,d) kagome lattice of $N_1\times N_2=8\times8$ unit cells. The data are shown along specific paths in the BZ. The interaction includes nearest-neighbour and next-nearest-neighbour repulsion $V_{\textrm{NN}}\sum_{\langle i,j \rangle}n_i n_j+V_{\textrm{NNN}}\sum_{\langle\langle i,j \rangle\rangle}n_i n_j$, with $V_\textrm{NN}=1$ and $V_\textrm{NNN}=0.2$. (a) The two-particle spectrum and (b) single-hole dispersion in the checkerboard lattice with the parameters $t_1=1,t_2=0.3,\phi=\pi/4$. The interaction is projected to the flattened $C=-1$ band. (c) The two-particle spectrum and (d) single-hole dispersion in the kagome lattice with the parameters $t_1=-1,\lambda_1=0.9$. The interaction is projected to the flattened $C=1$ band.
}
\label{2p1h}
\end{figure}

As an example, we first consider the two-particle spectrum of nearest-neighbour and next-nearest-neighbour repulsion projected to the flattened $C=-1$ band on a checkerboard lattice in Fig.~\ref{2p1h}(a). The spectrum depends on the total momentum $\mathbf K$, underscoring the lack of translation invariance in reciprocal space (in the Landau levels the corresponding eigenvalues are independent of the center of mass motion). There are only six finite levels, forming three approximately degenerate pairs (for the appropriate tight-binding parameters used in this example). By direct analogy to the two-particle spectrum in the lowest Landau level, these pairs can be labeled as $\mathcal V_1$, $\mathcal V_3$ and $\mathcal V_5$ from top to bottom, respectively. The separation of energy scales $\mathcal V_1\gg \mathcal V_3 \gg \mathcal V_5$ terms suggests stable FCIs above $\nu\gtrsim1/7$; for $\nu\geq 1/3$ they should be stabilized by the large energy scale, $\sim\mathcal V_1$, in the range $1/5\leq\nu<1/3$ the gap scale is set by $\mathcal V_3$ and for  $1/7\leq\nu<1/5$ the gaps are tiny, at the order of $\mathcal V_5$. This is consistent with our numerical observations for the particular model. In the original work of Ref.\cite{andreas}, it was shown by large scale exact diagonalization studies that a similar prediction for the nearest neighbor interaction on the checkerboard holds true; including solid evidence for the existence of a plethora of FCIs above $\nu=1/5$.

Depending on the tight binding parameters, the two-particle spectrum does not always look as nice and suggestive as in Fig.~\ref{2p1h}(a). Typically, the pairs of energy values are split and the dispersion as a function of the center of mass momentum, $\bb K$, can be significant. An example thereof is given by the two-particle dispersion of the nearest and next-nearest neighbor interactions projected to the lowest flat band of the kagome lattice model, as shown in Fig.~\ref{2p1h}(c). In this context there appears to be a genuine difference in the lattice between FCIs stabilized by interactions that mimic the corresponding FQH pseudopotentials, and those that do not have exact parent Hamiltonians in the continuum, such as the composite fermion\cite{jain} states (or more generally the hierarchy states\cite{haldane83,halperin84}). The states that are genuine zero modes of special pseudopotential Hamiltonians are not particularly sensitive to the precise details of the two-particle spectrum as long as there is an appropriate separation of energy scales [as is also the case in the kagome spectrum of Fig.~\ref{2p1h}(c)]. In contrast, the states that are not of this kind tend to be more sensitive to the details. In our two examples of the checkerboard and kagome lattice models, this is consistent with the observations of many hierarchy states in the former\cite{andreas} while certain model states are particularly stable in the latter.\cite{nonab3}

\subsection{Deviations from Landau level physics and new competing phases}\label{deviations}

Within a Landau level, any translation invariant two-body interaction is particle-hole symmetric. However, multi-body interactions---either in the form of psedopotential parent Hamiltonians or originating from perturbative corrections due to Landau level mixing---break this symmetry. In fact, at half-filling in the second Landau level this symmetry breaking is believed to be crucial for the low-energy physics favoring either the Moore-Read state\cite{mr} or its particle-hole conjugate, the so-called anti-Pfaffian state.\cite{apf1,apf2}

In Chern bands, the effect of particle-hole symmetry breaking is in fact even more pronounced and occurs already at the level of the projected two-body interaction.\cite{andreas} This is a direct consequence of the lack of translation invariance in the band, and is readily seen by performing a particle-hole transformation, $d_{\mathbf k,s} \rightarrow d_{\mathbf k,s}^\dagger$ within the band. Focusing on fermions, the projected Hamiltonian transforms to
\begin{eqnarray}H\rightarrow \sum_{\mathbf k_1,
 \mathbf k_2, \mathbf k_3, \mathbf k_4}
  (V^{{\rm proj}, s}_{\mathbf k_1 \mathbf k_2 \mathbf k_3 \mathbf k_4})^*d^{\dagger}_{\mathbf k_1,s}d^{\dagger}_{\mathbf k_2,s}d_{\mathbf k_3,s}d_{\mathbf k_4,s} +\sum_{\mathbf k}E_h(\mathbf k)
d^{\dagger}_{\mathbf k,s}d_{\mathbf k,s},\label{ph}\end{eqnarray}
which includes an effective single-hole energy
\begin{eqnarray}E_h(\mathbf k)=\sum_{\mathbf m}(
 V^{{\rm proj}, s}_{\mathbf m \mathbf k \mathbf m \mathbf k}+ V^{{\rm proj}, s}_{\mathbf k \mathbf m \mathbf k \mathbf m}- V^{{\rm proj}, s}_{\mathbf k \mathbf m \mathbf m \mathbf k}- V^{{\rm proj}, s}_{\mathbf m \mathbf k \mathbf k \mathbf m})\ .\end{eqnarray}
Within a Landau level, $E_h(\mathbf k)$ is simply a constant shift (chemical potential) while $E_h(\mathbf k)$ is generically dispersive in a Chern band due to the breaking of translation invariance in reciprocal space [cf. Figs.~\ref{2p1h}(b,d)]. It is crucial to note that this effective hole-dispersion arises in a completely flat band and is entirely induced by the collective behavior (interactions) and not due to a remnant kinetic term.\cite{andreas}

 \begin{figure}[bt]
\centerline{\psfig{file=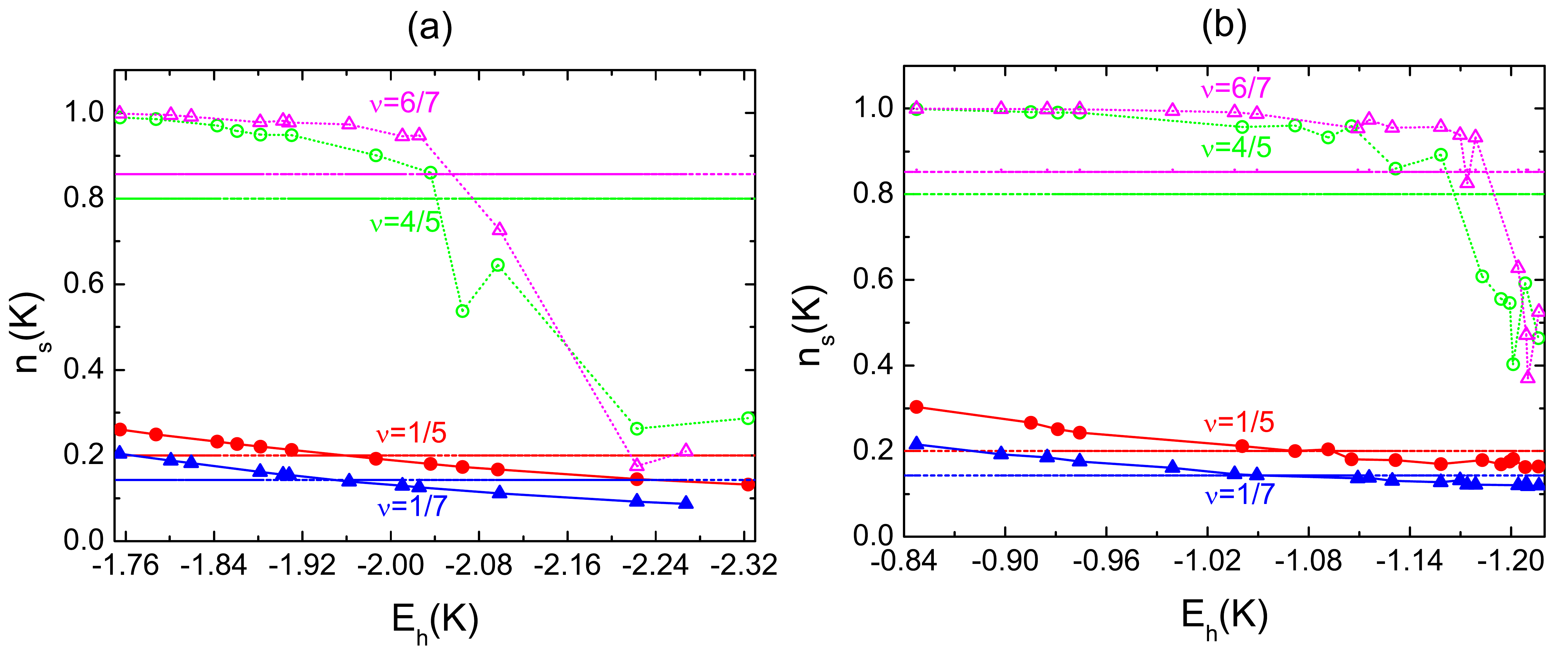,width=\linewidth}}
\vspace*{0pt}
\caption{\textbf{Illustration of the compressible states with an emergent Fermi surface at high filling fractions} in (a) checkerboard lattice and (b) kagome lattice. The plots show the behavior of $n_s(\mathbf k)$ as a function of the single-hole dispersion, $E_h(\mathbf k)$. The data are obtained by using the same interaction and tight-binding parameters as in Fig.~\ref{2p1h}. The numerical simulation is done for electrons in $N_1\times N_2=5\times6$ lattices at $\nu=1/5$ and $\nu=4/5$, and in $N_1\times N_2=5\times7$ lattices at $\nu=1/7$ and $\nu=6/7$. The dashed horizontal lines indicate the constant reference occupation $\av{n_s(\mathbf k)}=\nu$. The $\nu=4/5$ and $\nu=6/7$ states are very likely compressible because of the clearly visible Fermi-surface like feature in $n_s(\mathbf k)$ when plotted versus $E_h(\mathbf k)$.}
\label{nk}
\end{figure}

The non-constant hole dispersion leads to the absence of the particle-hole conjugate states of some stable low-filling FCIs. For example, FCIs are not observed at $\nu=4/5$ or $\nu=6/7$, even though they clearly exist at $\nu=1/5$ and $\nu=1/7$. At large filling fractions, the hole dispersion becomes dominant compared with the more conventional (two-hole) interaction term in Eq.~(\ref{ph}), and the ground states are expected to be compressible. In order to see this more clearly, we consider the momentum space occupation $n_s(\mathbf k)\equiv\langle d^{\dagger}_{\mathbf k,s}d_{\mathbf k,s}\rangle$ as a function of the hole dispersion $E_h(\mathbf k)$ (Fig.~\ref{nk}). At low filling factors, $n_s(\mathbf k)$ only slightly tracks $E_h(\mathbf k)$ with a roughly linear decrease, without deviating too much from the constant occupation $\nu$. This feature is perhaps somewhat unexpected, but is nevertheless completely consistent with an incompressible phase. However, as the filling fraction increases, the shape of $n_s(\mathbf k)$ is qualitatively distorted by $E_h(\mathbf k)$ and finally the Fermi-surface like structure appears, namely, $n_s(\mathbf k)\approx1$ for large $E_h(\mathbf k)$ but then markedly drops at some small $E_h(\mathbf k)$. This Fermi-surface like structure strongly suggests that the ground states are compressible Fermi-liquid like states (albeit supporting a non-quantized Hall effect). Large-scale exact diagonalizations for the checkerboard lattice model with the nearest neighbor repulsion demonstrate that FCIs are absent at $\nu\gtrsim2/3$.\cite{andreas} The similar behaviors in the kagome and checkerboard lattice models displayed in Fig.~\ref{nk} confirm that this instability is generic to the lattice case rather than an artifact of a specific flat band model.

As observed in Ref.\cite{Grushin}, an \emph{ad hoc} addition of a single particle dispersion can in principe restore particle-hole symmetry in the Chern band. Although this can in principle help stabilizing further FCIs,\cite{Grushin} it typically weaken, or even destroy, the FCIs observed at low filling fractions.\cite{AdiabaticContinuity3}

Charge density waves (CDWs) is another class of important competing phases. In particular, these are competitive when the interactions are relatively strong at the same time as the combination of the filling fraction and lattice geometry allows for favorable commensurate CDWs.\cite{ifwlong} A further deviation compared to the Landau level case is that the physics is typically more sensitive to changes in the aspect ratio in Chern bands.\cite{ThinTorusFCI} This is particularly clear when local interactions are considered, as the lattice comes with a minimal length scale, namely the lattice constant, while there is no such minimal length in the continuum (the magnetic length plays a somewhat different role there). For numerical investigations of interacting Chern band models it is therefore often desirable to focus on samples near unit aspect ratio.

\subsection{What characterizes a "good" flat band model?}\label{good}

The most obvious feature of a good flat band model is a large flatness ratio, and an interaction strength in a window such that Eq.~(\ref{projectioncriteria}) is fulfilled. However, in analogy with the FQH effect in conventional semiconductor heterostructures, FCIs might prevail far beyond this naive limit. In fact, in the conventional FQH setting, Eq.~(\ref{projectioncriteria}) is not fulfilled as the characteristic interaction strength, $V\sim e^2/(\epsilon\ell)$, is in fact typically similar to, or larger than, the cyclotron (band) gap, $\Delta=\hbar\omega_c$, while FQH states are nevertheless unambiguously observed\footnote{For instance, the experimental investigation of FQH states in the second Landau level reported in Ref.\cite{pan52} has $[e^2/(\epsilon\ell)]/\hbar \omega_c \approx 1.4$ in the vicinity of the $\nu=5/2$ plateau.}. There is also solid evidence confirming that FCIs can be stabilized in lattice models also when the interaction strength is substantially larger than the band gap, see e.g., Ref.\cite{cherninsnum1}. In fact, for the states that are zero modes of the interaction, the ratio $V/\Delta$ can be arbitrary large yet the band projection is (nearly) perfect as is the case for the FCI phases of hard-core bosons. For instance, for $\nu=1/2$ bosons in the lowest band of the Kapit-Mueller model,\cite{kapit}
the band projection is exact even for an infinite on-site repulsion in the limit that the exponential tail of arbitrary long-range hopping terms is taken into account. When the hopping is truncated, e.g., at the next-nearest-neighbor, the band projection is still excellent for any strength of the (on-site) interaction.

A second criterion often mentioned in the literature is a smooth Berry curvature. While this might often serve as a rule of thumb for a given lattice model, it can also lead to somewhat misleading conclusions. Again, this is most transparent in the Kapit-Mueller model\cite{kapit} where the Berry curvature is strongly varying\cite{eliotzhao}, but the model states [Laughlin ($k=1$), Moore-Read ($k=2$), Read-Rezayi (generic $k$)] are nevertheless exact gapped ground states for bosons with on-site $(k+1)-$body interactions. Nevertheless, if a close similarity to the full phase diagram of Landau level as a function of the filling fraction is the goal, a nearly constant Berry curvature and Fubini-Study metric\cite{roy,otherreview} often serve as useful indicators.

Furthermore, the details of the actual lattice interaction are crucial for which FCIs are realized. An often useful diagnostic for a given flat band model is to analyze the two-body spectrum\cite{andreas} and its relation to the pseudopotentials as explained in Section \ref{pseudop}. Most FCIs have indeed been found for short-range lattice interactions that closely mimic the respective continuum pseudopotentials, and long-range interactions, albeit potentially crucial in actual realizations, have been reported to weaken or destroy these phases. However, an upshot is that new phases may instead be stabilized in presence of longer-range interactions, in close analogy with the effectively longer-range interactions occurring in higher Landau levels where non-Abelian states might eventually be stabilized by very realistic two-body interactions (in contrast to the not particularly realistic multi-body nature of their respective pseudopotential parent Hamiltonians)\cite{eliotzhao}.

Another factor to take into account, especially for strong interactions comparable to the band gap, is whether or not there is a commensurate charge density wave (CDW) competing with the FCI phase.\cite{ifwlong} The existence of such competing states crucially depend on the combination of lattice geometry and the filling fraction. Finally, on general grounds we expect that a not too strong single-hole dispersion is desirable to avoid the gapless, albeit entirely interaction induced, states\cite{andreas} discussed in Section \ref{deviations}.

Summarizing, while there is no known single general diagnostic that immediately tells us how "good" a flat band model is, a combination of the flatness ratio, the two-particle spectrum, the single-hole dispersion, and when appropriate, also a consideration of possible competing CDW phases, typically gives a quite accurate idea of which FCI phases can be realized.

\section{Higher-$C$ models and new collective states of matter}\label{Cg1}

While the flat bands with $|C|=1$ mimic Landau levels, lattice models are known to host bands with arbitrary integer Chern number, $C\in \mathbb Z$. In this section we will first show that very flat bands with $|C|>1$ can also be achieved with short-range hopping terms\footnote{Note that given that dispersive $|C|>1$ models were known for a long time, Eq. (\ref{flattening}) would have provided a somewhat trivial but direct path to flat $|C|>1$ bands. However, a head-on application of this procedure leads to models that are necessarily long-ranged in the sense that they cannot be truncated to include only a few nearest neighbors while leaving the Chern number invariant.\cite{masa,longrange}} and then go on to discuss what is known about interacting phases in these bands, including FCIs with no direct continuum analogues. Finally, we  briefly comment on the prospects of finding new non-Abelian types of excitations bound to lattice dislocations in the $|C|>1$ FCIs.

\subsection{Flat band models with $|C|>1$}\label{flatbandCN}

Historically, the first higher $C$ model with flat bands proposed was a $C=2$ model on the dice lattice.\cite{c2} Next, a $C=2$ model was constructed on the triangular lattice.\cite{ChernTwo} A systematic construction of flat bands with arbitrary Chern number was first given in Ref.\cite{max} (see the example below). The spirit of this approach is easy to appreciate: given a system of $N$ decoupled layers, each supporting a flat $C=1$ band, it is possible to couple them in a way such that the $N$ initially degenerate bands hybridize to form a single topological band with $C=N$ while the others go trivial ($C=0$). In Ref.\cite{max} this was realized on a slab of the frustrated pyrochlore lattice which is composed of corner-sharing tetrahedra.
 In a subsequent publication\cite{dassarma} it was shown that the geometric frustration could be traded against the combination of inclusion of periodic hopping terms between top bottom layers\footnote{Alternatively this model can be re-written in terms of $N$ completely decoupled layers.\cite{private}} (and assigning different phase factors to the hopping processes in each layer).

\subsubsection*{Example: Pyrochlore slab model}

\begin{figure}[bt]
\centerline{\psfig{file=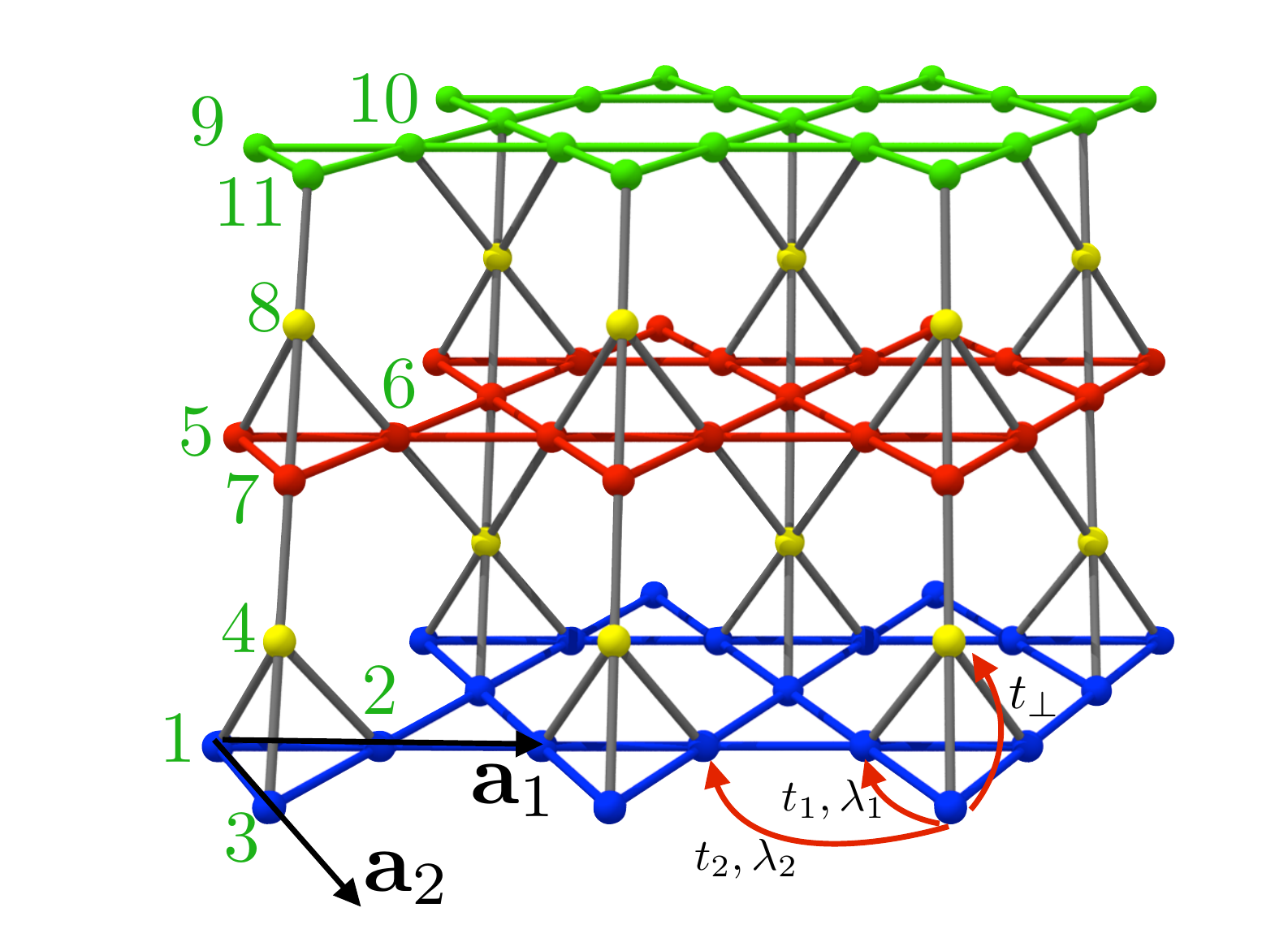,width=0.6\linewidth}}
\vspace*{8pt}
\caption{{\bf The spin-orbit coupled pyrochlore slab model}, here including $N=3$ kagome layers which are colored differently for clarity. The kagome layers are connected via sites living on $N-1$ intermediate triangular layers (sites of which are colored yellow). In general, the unit cell has $N_c=4N-1$ sites (green numbers) and the Bravais lattice is generated by the lattice vectors $\mathbf a_1$  and $\mathbf a_2$. The red arrows indicate considered nearest and next-nearest hopping processes, for which the signs of the spin-orbit terms, $\lambda_1,\lambda_2$, depend on the orientation of the process within a given hexagon in the pertinent kagome layer precisely as in the single layer case discussed in Section \ref{nonint}. The Bloch Hamltonian corresponding to the full pyrochlore slab tight-binding model is given by Eq.~(\ref{multilayerham}).
}\label{pyrochloreslab}
\end{figure}

Here we focus on the pyrochlore slab model\cite{max} (Fig.~\ref{pyrochloreslab}) and describe it in some detail. In addition to being the first flat band model with arbitrary $C$, it has a highly intriguing structure including a nontrivial relation between layer localization and momentum\cite{max} and it harbors novel forms of FCIs with no continuum analogue.\cite{ChernN,ChernN2}

For an arbitrary number of layers, $N$, the $(4N-1) \times (4N-1)$ Bloch Hamiltonian reads
\begin{align}
    \mathcal{H}^{N}_\bb k =& \begin{pmatrix}
        \mathcal{H}^{N=1}_\bb k & \mathcal{H}_{\perp,a} &  0 & 0 & 0 & 0\\
        \mathcal{H}_{\perp,a}^\dagger & 0  & \mathcal{H}_{\perp,b}^\dagger & 0 & 0& 0\\
        0 & \mathcal{H}_{\perp,b} & \mathcal{H}^{N=1}_\bb k  & \dots & 0 & 0\\
        0 & 0 & \vdots &  \ddots & \mathcal{H}_{\perp,a} & 0 \\
        0 & 0 & 0 &  \mathcal{H}_{\perp,a}^\dagger & 0 & \mathcal{H}_{\perp,b}^\dagger \\
        0 & 0 & 0 & 0 &\mathcal{H}_{\perp,b} & \mathcal{H}^{N=1}_\bb k \\
    \end{pmatrix} \ ,
    \label{multilayerham}
\end{align}
where $\mathcal{H}^{N=1}_\bb k$ is the Bloch Hamiltonian of a single kagome layer given in Eq.~(\ref{blochham}) and
$\mathcal{H}_{\perp,a}, \mathcal{H}_{\perp,b}$ encode the hopping processes to the triangular layers as
\begin{align}
\mathcal{H}_{\perp,a}
                = t_\perp \begin{pmatrix} 1 \\ 1 \\ 1 \end{pmatrix},\ \ \mathcal{H}_{\perp,b}= t_\perp \begin{pmatrix} e^{i k_2} \\ e^{-i k_3} \\ 1\end{pmatrix} \ .
\end{align}

\begin{figure}[bt]
\centerline{\psfig{file=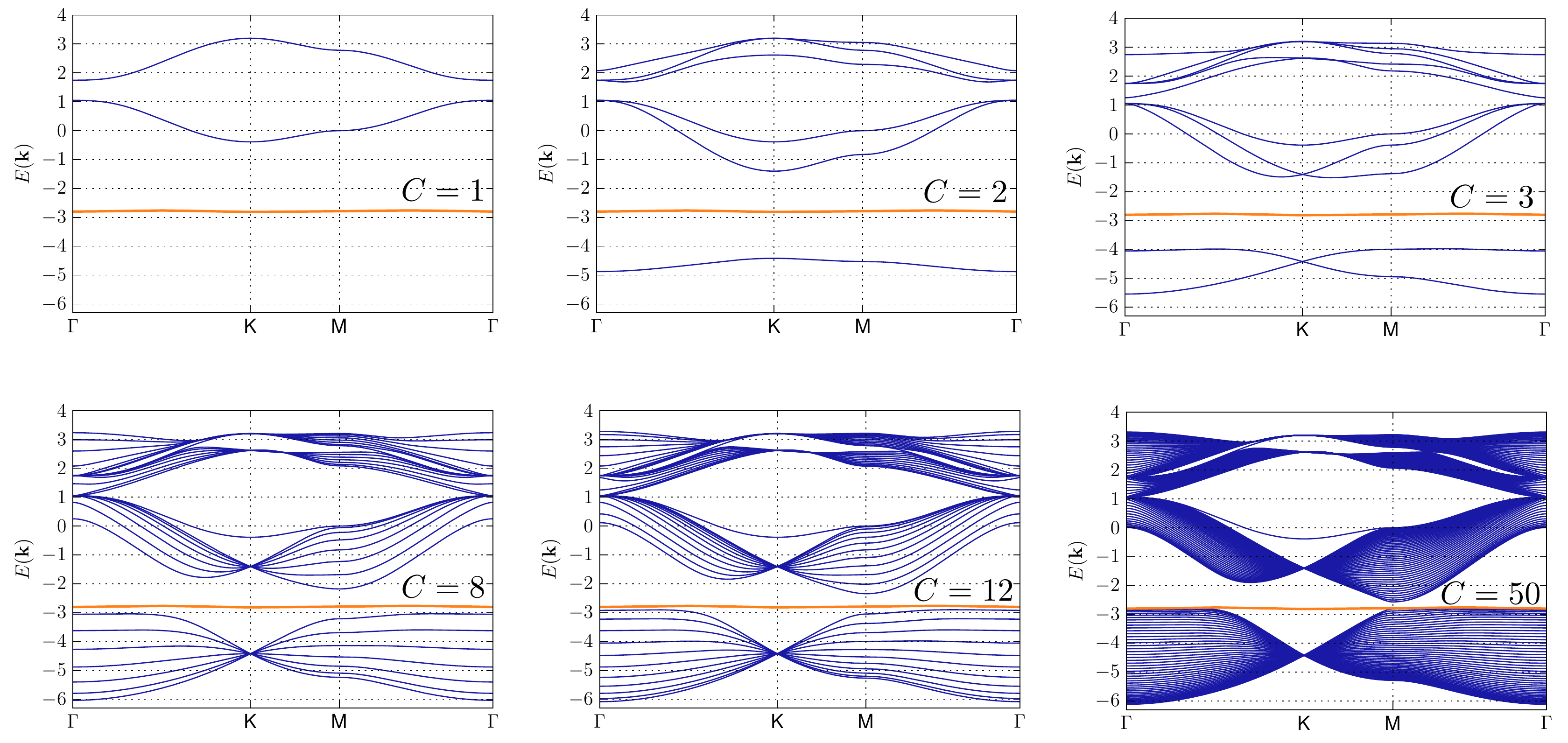,width=1.0\linewidth}}
\vspace*{8pt}
\caption{{\bf Band structure of the pyrochlore slab model}. Bulk dispersion for a system with $N=1,2,3,8,12,50$ stacked kagome layers, respectively. In each case there is a very flat band (bold orange line) with Chern number $C=N$. The parameters are chosen as $t_1=-1,t_2=\lambda_1=0.3,\lambda_2= 0.2,t_{\perp}=1.3$. As described in the text, the value of $t_{\perp}$ is not affecting the flat band but can be tuned to move the other bands (thin blue). 
}
\label{highCdispersion}
\end{figure}

\begin{figure}[t!]
\centerline{
\includegraphics[width=0.9\linewidth]{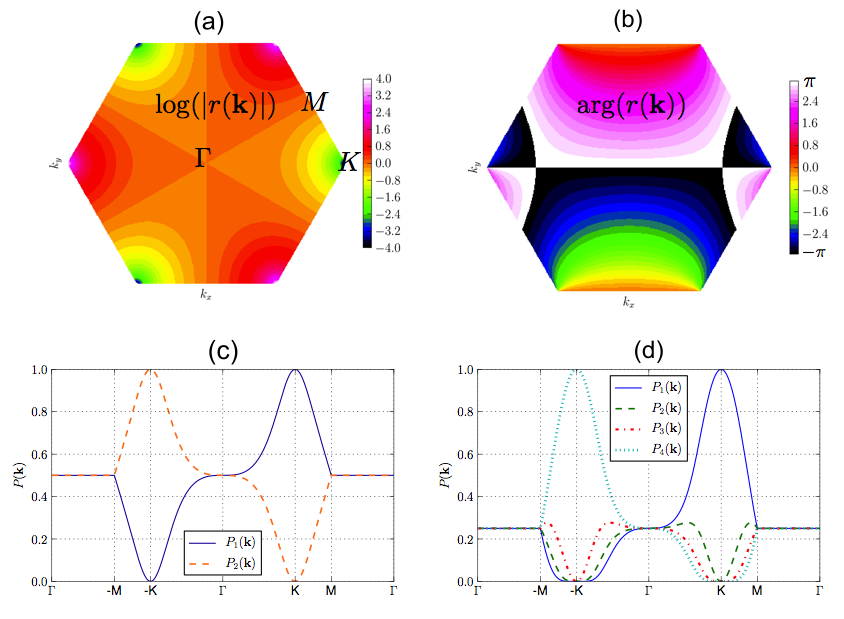}
}
\caption{{\bf Layer-momentum coupling in the high Chern band in the pyrochlore slab model}. (a) is a plot of $\log(|r(\mathbf k)|)$ for the flat band parameters $t_1=-1, t_2=0.3, t_\perp\neq 0, \lambda_1=0.3,\lambda_2=0.2$, which give a very flat $C=N$ band in the $N-$layer pyrochlore slab model. (b) shows the angle  $\arg(r(\mathbf k))$ for the same model and parameters. The lower panel (c),(d) show the resulting momentum dependent probability distribution, $P_m(\mathbf k)=|\alpha_m(\mathbf k)|^2$, to be within a given layer $m$ in the $C=N=2,4$ cases respectively. The complete localization to the top or bottom layers at the $K$-points corresponds to a divergence in $\log(|r(\mathbf k)|)$ and the equal layer population on the ray connecting the $M$ and $\Gamma$-points follows from $\log(|r(\mathbf k)|)=0$.
}
\label{fig:logrk}
\end{figure}

The bulk spectrum of the pyrochlore slab model, with a fixed set of tight binding parameters and various number of layers, $N$, is displayed in Fig.~\ref{highCdispersion}. The most striking feature is a very flat band (thick orange line) appearing at the same energy irrespective of the value of $N$. This strongly suggests an ansatz of the form
\begin{equation}
\ket{\psi_s(\mathbf k)}=\sum_{m=1}^{N}\alpha_m(\bb k)\ket{\phi_m^{C=1}(\mathbf k)}\label{ansatz}
\end{equation}
in terms of the single layer eigenstates $\ket{\phi_m^{C=1}(\mathbf k)}$.
It is clear that, to be of the form of Eq. (\ref{ansatz}), the eigenstates, $\ket{\psi_s(\mathbf k)}=\sum_b\psi^b_s(\bb k)c^\dagger_{{\bb k},b}\ket{0}$, have to have amplitudes of being on the triangular sites that vanish, i.e. \begin{equation}\psi^{4t}_{s}(\bb k)=0, \ \ t=1,\ldots, N-1\ .\end{equation}
Now, such states can only be eigenstates if the amplitudes of all hopping processes hopping to each of the sites in the triangular add up to zero\footnote{This is in complete analogy with how flat bands are built by localized states in usual nearest neighbor hopping problems (without spin-orbit coupling) on geometrically frustrated lattices.}. Using the basis indicated in Fig. \ref{pyrochloreslab}, this constraint implies
\begin{equation}
\psi^{4t-3}_{s}(\bb k)+\psi^{4t-2}_{s}(\bb k)+\psi^{4t-1}_{s}(\bb k)+e^{-ik_2}\psi^{4t+1}_{s}(\bb k)+e^{ik_3}\psi^{4t+2}_{s}(\bb k)+\psi^{4t+3}_{s}(\bb k)=0 \ .
\end{equation}
Together with the ansatz Eq. (\ref{ansatz}) this has a simple solution
\begin{align}
    \frac{\alpha_{m+1}(\bb{k})}{\alpha_m(\bb{k})} &= -\frac{\psi^{1}_{s}(\bb k)+\psi^{2}_{s}(\bb k)+\psi^{3}_{s}(\bb k)}{e^{-i k_2}\psi^{1}_{s}(\bb{k}) + e^{i k_3}\psi^{2}_{s}(\bb{k}) + \psi^{3}_{s}(\bb{k})}
    \equiv r(\bb{k}) \, .
    \label{alpharatio}
\end{align}
which uniquely determines the eigenstates in the flat band. By explicit calculation one may check that these states provide a band with $C=N$. It is a crucial fact that the coefficients, $\alpha_m(\bb k)$, are non-trivial in the sense that they depend on both $m$ and $\bb k$---if not the wave functions in Eq.~(\ref{ansatz}) would result in a $C=1$ band. Note however that the (complex) ratio $r(\bb{k}) = \frac{\alpha_{m+1}(\bb{k})}{\alpha_m(\bb{k})}$ does not depend on $m$, and that $r(\bb{k})$ can simply be extracted from the single-layer kagome model discussed in Section \ref{nonint}. This makes it easy to exhibit the intriguing structure of the eigenstates. Most saliently, for fixed $\bb k$ the states are exponentially localized, with a localization length given by $\log(|r(\bb k)|)$ to either the top or the bottom layer---which one is given by the sign of $\log(|r(\bb k)|)$ (cf. Fig. \ref{fig:logrk}). This intricate connection between layer index and momentum is somewhat reminiscent of the momentum spin correlation in time-reversal invariant topological insulators.\cite{toprev1}

Before closing this example we point out that it is useful to think about the band-structure as a function of $t_\perp$. As soon as $t_\perp$ is non-vanishing the bands hybridize; $N-1$ of them go trivial and become dispersive while a single band stays flat and acquires a finite Chern number, $C=N$, i.e., it 'absorbs all of the topology'. Although the analytical understanding of the structure of states in the $C=N$ bands crucially depended on the fact that the model was fine-tuned in the sense that the kagome layers are only coupled via a simple nearest neighbor hopping to the sites of the triangular layers, the gross features of the single particle states are very robust to the inclusion of more generic terms due to the topological nature of the band.

\subsection{Interactions and new bulk insulating states}\label{intercn}
Novel bulk insulating states are formed due to the crucial effect of interactions in the $C=N>1$ Chern bands. In Ref.\cite{ChernN}, a whole series of Abelian FCIs were discovered systematically at $\nu=1/(2N+1)$ for fermions and $\nu=1/(N+1)$ for bosons given local interactions in the pyrochlore slab model discussed in Section \ref{flatbandCN}. Moreover, a subsequent publication reported the observation of non-Abelian bosonic FCIs at $\nu=k/(N+1),k>1$ in the same model by introducing local $(k+1)$-body interactions.\cite{ChernN2} In other lattice models, FCIs that belong to the above series were also reported, partly independently, such as the $\nu=1/5$ fermionic state\cite{ChernTwo,Grushin,zhao2} and the $\nu=1/3$ bosonic state\footnote{The "composite fermion" states found earlier in an optical lattice setup with a uniform magnetic field\cite{gunnar} are close relatives of the $C=2$ FCIs. See also Ref.\cite{layla} for related results.}.\cite{dassarma,ChernTwo,zhao2}

\begin{figure}[t!]
\centerline{
\includegraphics[width=\linewidth]{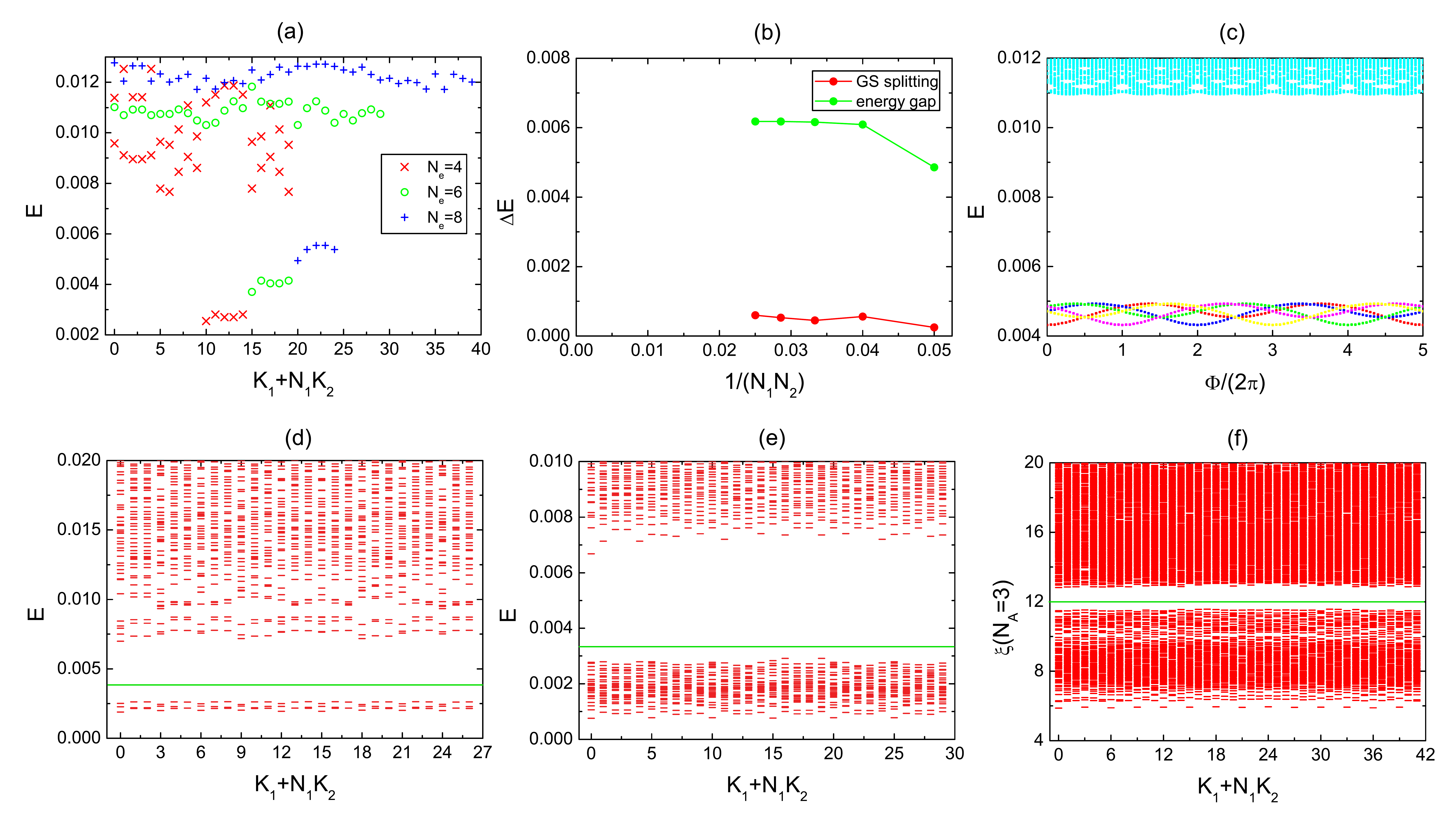}
}
\caption{\textbf{Numerical observation of FCIs in the $C=2$ band.} Results for the $\nu=1/5, C=2$ fermionic FCIs in a bilayer kagome system with
$N_1=5$ and $N_2=N_e$.
(a) The low-lying energy spectrum for $N_e=4$, $N_e=6$, and $N_e=8$.
(b) The finite-size scaling analysis for both energy gap and ground state splitting.
(c) The $x$-direction spectral flow for $N_e=7$.
(d) The quasihole excitations for $N_e=5$, $N_1=3$, and $N_2=9$ (two holes are added, 81 states below the gap).
(e) The quasihole excitations for $N_e=5$, $N_1=5$, and $N_2=6$ (five holes are added, 756 states below the gap).
(f) The particle-cut entanglement spectrum probing the $N_A=3$ quasihole excitations for $N_e=7$ (2695 states below the gap).
}
\label{fermion_1_5}
\end{figure}

As an example,\cite{ChernN} the numerical evidence of the $\nu=1/5$ fermionic FCIs in the $C=2$ band is shown in Fig.~\ref{fermion_1_5}. Here the band is flattened and the interaction is projected onto the flat band. A five-fold ground-state quasidegeneracy is observed for each system size, which is a necessary condition for the $\nu=1/5$ FCIs. The finite-size scaling analysis indicates that the gap, which is significantly larger than the ground state splitting, is very likely to survive in the thermodynamic limit. The evolution of states in the spectral flow suggests the Hall conductance $\sigma_{H}=\frac{2}{5}\frac{e^2}{h}$, which can also be confirmed by calculating the many-body Chern number. The total number of quasihole states of the system reflected by the quasihole excitation spectra and the particle-cut entanglement spectrum is the same as the one predicted by the generalized counting rule (no more than one electron in consecutive five 1D orbitals). All of the energetic and entanglement evidences above confirm the existence of the the $\nu=1/5$ fermionic FCIs in the $C=2$ flat band.

A quick glance at the results above will give a naive impression that the Abelian FCIs in higher Chern number bands are very similar to the spinless FQH Laughlin states in a single Landau level: they have the same topological degeneracy and the same total counting of quasihole states. However, closer inspection reveal some important differences: the counting in each momentum sector does not match that of the Laughlin states,\cite{ChernN2} and the total counting in the particle-cut entanglement spectrum is lower than expected in some $|C|>2$ cases,\cite{ChernN,ChernN2} which suggest a colorful explanation of these novel Abelian FCIs: they are related to the color-dependent magnetic-flux inserted version of Halperin states\footnote{It is suggested that those non-Abelian FCIs in higher Chern number bands can be related to color-dependent magnetic-flux inserted version of the non-Abelian spin-singlet states.\cite{BlochFCI}}.\cite{BlochFCI,ChernN2} Moreover, although e.g., $\nu=1/5$ is not a very surprising fraction for finding a fermionic FCI, it should be noted that there is no evidence that the $C=2$ bands harbor a fermionic FCI at $\nu=1/3$. Also, the discovered bosonic FCIs appear at anomalous (odd denominator) filling fractions in striking contrast to the Landau level case.

A very exiting suggestion\cite{disloc} due to Barkeshli and Qi is that lattice dislocations could act like 'wormholes' changing the geometry of space when appearing in $|C|>1$ bands hosting FCIs. Moreover, the dislocations would also carry a non-trivial quantum dimension and behave as non-Abelian excitations. A simple geometric picture captures the gist of the (field theory based) argumentation by thinking of the initial system as a multilayer system. Each 'layer' can then be thought of as a torus (periodic boundary conditions), and a dislocation connects different tori and thereby increases the genus, $g$, of space (the genus simply counts the number of topologically inequivalent loops). Now, since FQH states are known to have a degeneracy that depends on the genus of the underlying manifold, for the $\nu=1/q$ Laughlin states the degeneracy is $q^g$,\cite{wenniu} adding more dislocations increases the ground state degeneracy and makes it meaningful to assign a quantum dimension, $d>1$, which counts how quickly the ground state manifold grows as a excitations at fixed positions are added to the system, to each dislocation. We note that the many-body states assumed by Barkeshli and Qi were different from the states reported so-far in the numerical literature. However, by tuning the interactions one can indeed realize the starting point of Ref.\cite{disloc}.\cite{joergsam} Nevertheless, microscopic tests the  predictions in actual strongly correlated lattice models including dislocations are still lacking.

\section{Paths toward experimental realization}\label{exp}

Realizing the rich phenomenology offered by FCIs in actual experiments is arguably a most pressing issue. There are several more or less feasible routes towards achieving this. Below we give some brief comments on a few selected ideas divided into the two main directions, namely suggested solid state implementations and cold atom/molecule engineering.

\subsection{Solid state implementations} \label{solidstate}
The possibility of realizing FCIs in spin-orbit coupled solid state materials was discussed already in one of the initial works focusing on tentative kagome lattice candidates.\cite{chernins1} Significant steps towards realistic flat band models were taken in Refs.\cite{c2,digital,ifw1} focusing on transition-metal oxides. The quite realistic first principle calculations of Ref.\cite{digital} were particularly encouraging. Here, it was found that sandwich structures of certain perovskite materials grown in the $[111]$ direction, where buckled honeycomb structures would ideally form, can support relatively flat bands carrying $C=1$. Although there is an intense activity in the materials science community aiming to grow such heterostructures, a significant hurdle to overcome is that the $(111)$ planes are not natural cleavage planes, thus growing clean perovskites $[111]$ direction is very challenging. This, and resulting reconstruction processes, may explain why there is, to the best of our knowledge, no experimental realization of these topological flat bands even though interesting experimental progress, notably including epitaxial growth of ultra thin $(111)$-oriented LaAlO$_3$/LaNiO$_3$ superlattices, was recently reported.\cite{111perovskites} A natural alternative to the perovskites is to grow pyrochlore based transition-metal oxides along the $[111]$ direction.\cite{max,fiete} A major advantage compared to the $(111)$ perovskite slabs is that the pyrochlore structure naturally admits growth/cleavage along the $[111]$ direction (cf. Fig.~\ref{pyrochloreslab}). Moreover, this would, at least in principle, open the door towards an experimental realizations of exotic $|C|>1$ phenomena.\cite{max,disloc}

As an (integer) anomalous Hall effect, and hence the first Chern insulator (in absence of an external magnetic field), was experimentally reported very recently in thin films of chromium-doped (Bi,Sb)$_2$Te$_3$,\cite{Chernexp} similar systems may also be of interest in the pursuit for FCIs. To mention a few other solid state proposals, we note that recent materials considerations suggest that 'organometallic frameworks' as potential hosts of topological flat bands\cite{organometallic,organometallic2} and that strained graphene provides yet another suggested FCI platform.\cite{grapheneFCI}

\subsection{Cold atom and molecule engineering}\label{coldatom}
As often when some degree of fine tuning of parameters is desirable, implementations using cold atom systems offer intriguing possible routes towards implementing fractional Chern insulators in experiments. It should however be noted that the celebrated tunability in these systems is essentially always limited to the strength of the on-site interaction and, more crucially, typically to nearest neighbor hopping processes only. Nevertheless, recent ideas in the context of 'artificial gauge fields',\cite{artgauge} whereby the complex internal structure of the atoms interacting with externally controlled laser fields is used to simulate the effect of a magnetic field, undoubtedly provides a promising route towards possible realizations of FCIs.

Lately there are two more detailed suggestions that seem particularly interesting\cite{DipolarTFB2,opticalflux} (see also Ref.\cite{viewpoint2}).
The first approach is based on polar molecules and the FCI would not be a collective state involving itinerant molecules (they are pinned on their respective lattice position), rather their rotational degrees of freedom are suggested to form an FCI\cite{DipolarTFB2,DipolarTFB} (see Ref.\cite{taoshi} for a somewhat similar suggestion to realize topological flat band of phonons in trapped-ion systems). Neighboring molecules can exchange local rotational states, similar to a spin-flip, and while doing so crucially pick up a phase factor. Now, the 'spin flips'  behave like hard-core bosons and they form two bands with non-zero Chern numbers, $C=\pm 1$.\cite{DipolarTFB} At half-filling in the lower band, the interactions indeed lead to an FCI state with Laughlin-like character, which can be possibly detected and characterized by measuring the single spin-flip response of the system.\cite{DipolarTFB2}
In the second approach, the 'optical flux lattices', several laser beams are used to produce a spatially  inhomogenous (periodic) atom-laser coupling which induces resonant transitions between different internal atomic states and can be described as a tight-binding model in reciprocal space.\cite{coopermoessner} This gives flat bands with variable Chern numbers suitable for realizing various FCIs, a signal of the formation of which would be the appearance of density plateaus in \emph{in situ} images of the atom gas.\cite{opticalflux}

\section{Discussion and outlook}\label{discussion}

In this review we have provided a rather detailed account of the recent flurry of works on topological flat band models characterized by non-zero Chern numbers, as well as the crucial effect of interactions within these bands. Our focus has been on microscopic models and we have attempted to provide a holistic view on the interacting phase diagram by combining insights from numerical investigations coupled with analytical considerations, with an emphasis on the connection to continuum Landau levels and their FQH phases, as well as on new lattice specific instabilities and phenomena. Before closing we now wish to share some of our thoughts on a few directions that we believe deserve extra attention.

\subsubsection*{Competing phases}

Quite naturally, the initial focus in the study of topological flat band models has been on the incompressible phases, the FCIs. However, a broader understanding of the phase diagram is desirable for several reasons. It would deepen our understanding on the circumstances needed for realizing FCIs, and competing phases might represent new intriguing interesting phases of matter in their own right. Moreover, once topological flat bands are realized experimentally (as we hope and believe will happen) it might very well be that phases other than the FCIs are easier to realize.

Perhaps the most interesting FCI competitors are the interaction induced compressible states\cite{andreas} discussed in Section \ref{deviations}. Although a Fermi surface like feature is clearly visible as a function of the 'single-hole dispersion', the nature of these states are not that well explored. This appears to be a pressing question given that these states are generic to topological flat band models rather than features of certain models. Most saliently, it would be interesting to quantitatively understand how these phases would be manifested in transport measurements, and what instabilities the 'Fermi surfaces' allow. In transport, there is presumably a non-quantized anomalous Hall effect but the details thereof, including the response to external probes such as electric and magnetic fields, remain to be studied. A better understanding of the structure of Fermi surfaces (or pockets etc.) in reciprocal space would lead to new interesting questions such as: what is the the effect of residual interactions, lattice vibrations (phonons) and/or (proximity induced) superconductivity? It would also be interesting to investigate possible connections to earlier studies of Berry curvature effects in more conventional time reversal broken Fermi liquids\cite{tfl,tfl2} and to composite fermion Fermi liquids known from half-filling in the lowest Landau level.\cite{hlr}

Another interesting, but so-far scarcely studied, class of states are those that do not admit a single band description but are nevertheless topologically non-trivial (see Ref.\cite{cdw+fci} for an example). Further, it would be of interest to understand if new competing phases emerge in the limit of high Chern numbers.

\subsubsection*{Microscopic models for dislocations}
The predicted non-Abelian statistics of lattice dislocations,\cite{disloc} which are ubiquitous in solid state materials, is arguably one of the most exciting possible phenomena in topological flat band models beyond conventional FQH physics. On a fundamental level, two hurdles remain to establish this idea in microscopic models. First, the starting point for the analysis of Ref.\cite{disloc} assumes a type of bi-layer states that have not yet been found in the FCI literature. Preliminary results however show that this first hurdle can be vanquished.\cite{joergsam} More importantly, it is not clear to what extent short-distance physics arising due to the introduction of lattice defects (dislocations) in actual microscopic models might qualitatively modify the long-distance (field theoretic) considerations in Ref.\cite{disloc}. On a practical level, there is a definite need for more  detailed predictions where to find these phenomena (see however Section \ref{solidstate}).

\subsubsection*{New perspectives, quantum geometry and tensor networks }

In this review we have shown many examples of how well FQH model wave functions and pseudopotential analogies can be used to describe FCIs. However, albeit being very enlightening, certain new aspects of the physics beyond the FQH paradigm are somewhat hidden by this success. As emphasized lately by Haldane (see e.g., Ref.\cite{haldanegeometry}), the fundamental problem to solve in Chern bands is not involving nicely behaved analytic wave functions, as in ideal Landau levels, but rather the 'quantum geometric' problem of minimizing the mutual interaction between band projected density operators. One may therefore ask, are there solvable models beyond the realm of pseudopotentials, e.g., in models that do not (exclusively) yield translationally invariant ground states (in reciprocal space)? At the level of toy-models the answer is clearly yes as was shown in Ref.\cite{ttmps}, where a parameter family of models relevant for a ladder version of a topological flat band model, which does not require translation invariance, was solved exactly. It is an open problem how to extend this to more realistic, truly two-dimensional models. An interesting framework in this context was outlined in Ref.\cite{pairingproblem} and makes use of an intriguing connection to pairing problems which have a rich history in the context of superconductivity.

Another natural idea would be to use matrix product states (MPS) approaches. In fact, the ground states of the model studied in Ref.\cite{ttmps} have an exact MPS representation, and moreover, there has recently been significant progress in understanding the connection between MPS states and FQH states (including both the model states\cite{mpsfqh,mpsfqh2} and the ground states of generic two-body interactions\cite{mpsfqh3}). Indeed, recent works focusing on $|C|=1$ FCIs have demonstrated that MPS based numerical approaches can significantly extend the range of system sizes reachable by conventional exact (numerical) diagonalization techniques also for these systems.\cite{dmrg,cincio,balents} Clearly, extending this to $|C|>1$ models would be highly desirable.

Despite impressive progress with MPS representations, their one-dimensional nature implies a fundamental limitation due to the area law of entanglement entropy\footnote{The computational cost using MPS approaches scales exponentially in the linear width of the sample. This is much better than being exponential in both dimensions as in exact diagonalization approaches, but is still a fundamental limitation of the method when applied to two-dimensional samples.}.\cite{arealaw}
The tensor-network algorithms,\cite{tnreview,tnreview2} such as the projected entangled pair states (PEPS) or the multi-scale entanglement renormalization ansatz (MERA), are developed to in principle overcome the fundamental limitations of MPS approaches in higher dimensions, and may therefore offer intriguing possibilities for the simulation of truly large topological flat band systems. Although a naive estimate of the entanglement content (density) of FCIs suggests that this should be possible, there is no successful application thereof to this date. A common conception is that chiral phases, such as the FCIs, cannot be simulated efficiently using tensor networks for a seemingly fundamental reason: in general, the tensors can be used to construct the edge theory given any spatial cut of the ground state wave function, but at the same time the edge states are gapless and chiral which means that the tensor dimension would have to be infinite (moreover, the chiral edge states need a two-dimensional bulk to exist in the first place). Nevertheless, even if there are limitations obstructing the study of truly infinite systems, tensor network methods may still be a valuable tool, and lead us to a better understanding of the microscopics of the FCIs. Promisingly, when finalizing this we learned that the 'fundamental' hurdle might perhaps be possible to overcome, at least in some cases.\cite{tensorchiral,tensorchiral2} See also Ref.\cite{beri} for a precursor study.

Finally, we note that topological flat band models allow a local Hamiltonian with a finite-dimensional local Hilbert space, as they can be studied without invoking band projection. This is in sharp contrast to the conventional FQH setting, where band projection is crucial for formulating a lattice problem in the first place. With the band projection, the unavoidable algebraic tail of the wave functions would present a tremendous complication---thus, in this sense the topological flat band problem is much more amenable to numerical simulation than the Landau level problem.

\subsubsection*{Other types of fractional topological insulators}

Fractional Chern insulators are arguably the most natural strongly correlated phases to expect in topological flat band models due the rich history of the fractional quantum Hall effect. They are, however, not the only possibility of topologically ordered phases with concomitant new phenomena including fractionalized excitations, induced by strong interactions in systems with topologically non-trivial band structure. Let us comment on these possibilities in ascending dimensions (and excitement).

There has been a lot of recent activity aiming to understand the interplay between interactions and topology in various one-dimensional systems. We note that, contrary to recent claims that these systems harbor faithful analogues of FQH states, these strictly one-dimensional interacting phases are, at best, CDWs with symmetry protected nontrivial Berry phase, as clarified in Ref.\cite{jan}. Similarly, one-dimensional non-interacting quasi-crystals have been argued to possess higher-dimensional topological phases. However, these are also not topological in the conventional sense\cite{kevin}---rather, assuming no symmetries, only families of one-dimensional Hamiltonians, aka pumps, may faithfully represent topological phases in higher dimensions as is known since the work of Thouless.\cite{thoulesspump} Note that this does not contradict that a one-dimensional point of view can be very helpful for understanding various aspects of FQH phases.\cite{bklong,rootc,pattern,seidel}

The situation is more promising in two dimensions. Beyond the time reversal broken phases (such as the FCIs), symmetry protected fractional phases are easy to imagine. Levin and Stern\cite{FTI} noticed that FCIs can be used as building blocks for time-reversal symmetric topological insulators by combining copies of time reversed FQH/FCI states and formulated a simple criteria for which states of this type that can in principle topological in the sense that the edge modes are stable (see also Ref.\cite{qsh}). Despite very interesting investigations,\cite{neupertfti,neupertfti2} however, all time reversal symmetric fractional topological insulators found in microscopic models so far seem to be essentially simple combinations of FCIs. A possible twist on this approach would be to combine FCIs living in bands with variable $|C|$, e.g., as done at the non-interacting level in Ref.\cite{peng}. Moreover, that qualitatively new physics has not yet been uncovered in the studies of time reversal symmetric fractional topological insulators does by no means imply that it does not exist. Rather, in our view, it should be perceived as an encouragement for further studies of these systems.

If anything, the situation seems even more exciting in three spatial dimensions. A recently pointed out possibility is that flat surface bands of three-dimensional topological insulators give rise to new physics.\cite{flatsurface} Most saliently, it is in principle possible that the surface states, which have previously been believed to be necessarily gapless, can instead form gapped states of matter provided that they possess non-Abelian topological order.\cite{surfaceFQH,surfaceFQH2,surfaceFQH3,surfaceFQH4} Thus interactions can have profound implications on the physics of topological insulators also in three dimensions.\cite{3dintclass} Even more ambitiously, one may think about phases with bulk topological order in three-dimensional systems. Although ideas on the level of effective field theories exist in the literature,\cite{3dFTI,3dFTI2} and the fact that there exist tight-binding models with three-dimensional topological flat bands,\cite{3dflatbands} the demonstration of a topologically ordered fractional topological insulator in three spatial dimensions would represent a significant leap in the theory of quantum matter.

\section*{Acknowledgements}

We acknowledge useful discussions and related collaborations with J. Behrmann, P. Brouwer, G. Haack, E. Kapit, D. Kovrizhin, A. L\"auchli, R. Moessner, M. Nakamura, S. Sanchez, J. Suorsa, M. Trescher and M. Udagawa. We specially thank  J. Behrmann, D. Kovrizhin and S. Sanchez for valuable comments on the manuscript and to A. L\"auchli and R. Moessner for the collaboration on Ref.\cite{andreas} which was instrumental for our understanding of the topic of the present review. E.~J.~B. also thanks G. M\"oller, N. Regnault, S. Simon, K. Sun, Y-.L. Wu, for discussions that proved valuable when finalizing this work. E.~J.~B. was supported by the Alexander von Humboldt foundation and by DFG's Emmy Noether program (BE 5233/1-1). Z.~L.~is supported by the China Postdoctoral Science Foundation Grant No. 2012M520149.



\section*{References}

\begin{thebibliography}{0}

\bibitem{vonk} K.v. Klitzing, G. Dorda and M. Pepper, {\em New Method for High-Accuracy Determination of the Fine-Structure Constant Based on Quantized Hall Resistance},
    \href{http://link.aps.org/doi/10.1103/PhysRevLett.45.494}{Phys. Rev. Lett. {\bf 45}, 494 (1980)}.

\bibitem{tsui}  D.C. Tsui, H.L. St\"ormer and A.C. Gossard, {\em Two-Dimensional Magnetotransport in the Extreme Quantum Limit}, \href{http://link.aps.org/doi/10.1103/PhysRevLett.48.1559}{Phys. Rev. Lett. {\bf 48}, 1559 (1982)}.

\bibitem{toprev1}M. Z. Hasan and C. L. Kane,  {\em Colloquium: Topological insulators},
    \href{http://link.aps.org/doi/10.1103/RevModPhys.82.3045}{Rev. Mod. Phys. {\bf 82}, 3045 (2010)}.

\bibitem{toprev2}X.-L. Qi and S.-C. Zhang,  {\em Topological insulators and superconductors}, \href{http://link.aps.org/doi/10.1103/RevModPhys.83.1057}{Rev. Mod. Phys. {\bf 83}, 1057 (2011)}.

\bibitem{it1} R.B. Laughlin, {\em Quantized Hall conductivity in two dimensions},
    \href{http://link.aps.org/doi/10.1103/PhysRevB.23.5632}{Phys. Rev. B. {\bf 23}, 5632 (1981)}.

\bibitem{it2} B.I. Halperin, {\em Quantized Hall conductance, current-carrying edge states, and the existence of extended states in a two-dimensional disordered potential},
    \href{http://link.aps.org/doi/10.1103/PhysRevB.25.2185}{Phys. Rev. B. {\bf 25}, 2185 (1982)}.

\bibitem{tknn}D. J. Thouless, M. Kohmoto, M. P. Nightingale, and M. den Nijs,  {\em Quantized Hall Conductance in a Two-Dimensional Periodic Potential},
    \href{http://link.aps.org/doi/10.1103/PhysRevLett.49.405}{Phys. Rev. Lett. {\bf 49}, 405 (1982)}.

\bibitem{avron}J. E. Avron, R. Seiler, and B. Simon, {\em Homotopy and quantization in condensed matter physics}, \href{http://link.aps.org/doi/10.1103/PhysRevLett.51.51}{Phys. Rev. Lett. {\bf 51}, 51 (1983)}.

\bibitem{qniu}Q.~Niu, D.~J.~Thouless, and Y.-S.~Wu, {\em Quantized Hall conductance as a topological invariant}, \href{http://link.aps.org/doi/10.1103/PhysRevB.31.3372}{Phys. Rev. B {\bf 31}, 3372 (1985)}.

\bibitem{resdef}P. J. Mohr, B. N. Taylor, and D. B. Newell, {\em CODATA recommended values of the fundamental physical constants: 2006}, \href{http://link.aps.org/doi/10.1103/RevModPhys.80.633}{Rev. Mod. Phys. {\bf 80}, 633 (2008)}.

\bibitem{Laughlin83} R.B. Laughlin,  {\em Anomalous Quantum Hall Effect: An Incompressible Quantum Fluid with Fractionally Charged Excitations},  \href{http://link.aps.org/doi/10.1103/PhysRevLett.50.1395}{Phys. Rev. Lett. {\bf 50}, 1395 (1983)}.

\bibitem{toporder}X.-G. Wen, {\em Topological Orders in Rigid States},
    \href{http://dx.doi.org/10.1142/S0217979290000139}{Int. J. Mod. Phys. B {\bf 4}, 239 (1990)}.

\bibitem{wenbook}X.~G.~Wen, {\em Quantum Field Theory of Many-body Systems}, (Oxford Graduate Texts, 2004).

\bibitem{wenniu} X.-G. Wen and Q. Niu, {\em Ground state degeneracy of the FQH states in presence of random potential and on high genus Riemann surfaces},
    \href{http://link.aps.org/doi/10.1103/PhysRevB.41.9377}{Phys. Rev. B {\bf 41}, 9377 (1990)}.

\bibitem{kitaevpreskill} A.~Kitaev and J.~Preskill, {\em Topological Entanglement Entropy}, \href{http://link.aps.org/doi/10.1103/PhysRevLett.96.110404}{Phys.~Rev.~Lett. {\bf 96}, 110404 (2006)}.

\bibitem{levinwen} M.~Levin and X.~G.~Wen, {\em Detecting Topological Order in a Ground State Wave Function}, \href{http://link.aps.org/doi/10.1103/PhysRevLett.96.110405}{Phys.\ Rev.\ Lett.\ {\bf 96}, 110405 (2006)}.

\bibitem{mr} G. Moore, and N. Read, {\em Nonabelions in the fractional quantum Hall effect},
    \href{http://dx.doi.org/10.1016/0550-3213(91)90407-O}{Nucl. Phys. B {\bf 360}, 362 (1991)}.

\bibitem{topocomp}
C. Nayak, S.H. Simon, A. Stern, M. Freedman, and S. Das Sarma, {\em Non-Abelian anyons and topological quantum computation},
    \href{http://link.aps.org/doi/10.1103/RevModPhys.80.1083}{Rev. Mod. Phys. {\bf 80}, 1083 (2008)}.

\bibitem{fstat}J. M. Leinaas and J. Myrheim, {\em On the theory of identical particles},
    \href{http://dx.doi.org/10.1007/BF02727953}{Nuovo Cimento Soc. Ital. Fis. {\bf 37B}, 1 (1977)}.

\bibitem{majoranareturns}See e.g. F. Wilczek, {\em Majorana returns}, \href{http://dx.doi.org/10.1038/nphys1380}{Nat. Phys. {\bf 5}, 614  (2009)}, for a brief history of the subject.

\bibitem{willett}R. L. Willett, L. N. Pfeiffer, and K. W. West, {\em Alternation and interchange of $e/4$ and $e/2$ period interference oscillations consistent with filling factor $5/2$ non-Abelian quasiparticles},
    \href{http://link.aps.org/doi/10.1103/PhysRevB.82.205301}{Phys. Rev. B {\bf 82}, 205301 (2010)}.

\bibitem{mourik} V. Mourik, K. Zuo, S.M. Frolov, S.R. Plissard, E.P.A.M.
Bakkers, and L.P. Kouwenhoven, {\em Signatures of Majorana Fermions in Hybrid Superconductor-Semiconductor Nanowire Devices},
    \href{http://www.sciencemag.org/content/336/6084/1003.abstract}{Science {\bf 336}, 1003 (2012)}.

\bibitem{haldanemodel} F.D.M. Haldane,
{\em Model for a Quantum Hall Effect without Landau Levels: Condensed-Matter Realization of the "Parity Anomaly"},
    \href{http://link.aps.org/doi/10.1103/PhysRevLett.61.2015}{Phys. Rev. Lett. {\bf 61}, 2015 (1988)}.

\bibitem{kanemele}C. L. Kane and E. J. Mele, {\em Quantum Spin Hall Effect in Graphene},
    \href{http://link.aps.org/doi/10.1103/PhysRevLett.95.226801}{Phys. Rev. Lett. {\bf 95}, 226801 (2005)}.

\bibitem{kanemele2}C. L. Kane and E. J. Mele , {\em $Z_2$ Topological Order and the Quantum Spin Hall Effect}, \href{http://link.aps.org/doi/10.1103/PhysRevLett.95.146802}{Phys. Rev. Lett. {\bf 95}, 146802 (2005)}.


\bibitem{Hofstadter}D. R. Hofstadter, {\em Energy levels and wave functions of Bloch electrons in rational and irrational magnetic fields}, \href{http://link.aps.org/doi/10.1103/PhysRevB.14.2239}{Phys. Rev. B {\bf 14}, 2239 (1976)}.

\bibitem{viewpoint1} R. Roy and S. L. Sondhi, {\em Fractional Quantum Hall Effect without Landau Levels}, \href{http://link.aps.org/doi/10.1103/Physics.4.46}{Physics {\bf 4}, 46 (2011)}.

\bibitem{kapit}E.~Kapit and E.~Mueller, {\em Exact Parent Hamiltonian for the Quantum Hall States in a Optical Lattice}, \href{http://link.aps.org/doi/10.1103/PhysRevLett.105.215303}{Phys.~Rev.~Lett.~{\bf 105}, 215303 (2010)}.

\bibitem{chernins1}
E. Tang, J.-W. Mei, and X.-G. Wen, {\em High-Temperature Fractional Quantum Hall States},
    \href{http://link.aps.org/doi/10.1103/PhysRevLett.106.236802}{Phys. Rev. Lett. {\bf 106}, 236802 (2011)}.

\bibitem{chernins2}
K. Sun, Z. Gu, H. Katsura, and S. Das Sarma, {\em Nearly Flatbands with Nontrivial Topology},
    \href{http://link.aps.org/doi/10.1103/PhysRevLett.106.236803}{Phys. Rev. Lett. {\bf 106}, 236803 (2011)}.

\bibitem{chernins3}
T. Neupert, L. Santos, C. Chamon, and C. Mudry, {\em Fractional Quantum Hall States at Zero Magnetic Field},
    \href{http://link.aps.org/doi/10.1103/PhysRevLett.106.236804}{Phys. Rev. Lett. {\bf 106}, 236804 (2011)}.

\bibitem{cherninsnum1}
D. N. Sheng, Z. Gu, K. Sun, and L. Sheng, {\em Fractional quantum Hall effect in the absence of Landau levels},
    \href{http://dx.doi.org/10.1038/ncomms1380}{Nat. Commun. {\bf 2}, 389 (2011)}.

\bibitem{cherninsnum2}
N. Regnault and B. A. Bernevig, {\em Fractional Chern Insulator},
    \href{http://link.aps.org/doi/10.1103/PhysRevX.1.021014}{Phys. Rev. X {\bf 1}, 021014 (2011)}.

\bibitem{bosons}
Y.-F. Wang, Z.-C. Gu, C.-D. Gong, and D. N. Sheng, {\em Fractional Quantum Hall Effect of Hard-Core Bosons in Topological Flat Bands},
    \href{http://link.aps.org/doi/10.1103/PhysRevLett.107.146803}{Phys. Rev. Lett. {\bf 107}, 146803 (2011)}.

\bibitem{c2}
F. Wang and Y. Ran,
{\em Nearly flat band with Chern number $C=2$ on the dice lattice},
    \href{http://link.aps.org/doi/10.1103/PhysRevB.84.241103}{Phys. Rev. B {\bf 84}, 241103(R) (2011)}.

\bibitem{max}
M. Trescher and E. J. Bergholtz, {\em Flat bands with higher Chern number in pyrochlore slabs},
    \href{http://link.aps.org/doi/10.1103/PhysRevB.86.241111}{Phys. Rev. B {\bf 86}, 241111(R) (2012)}.

\bibitem{dassarma} S. Yang, Z.-C. Gu, K. Sun, and S. Das Sarma, {\em Topological flat band models with arbitrary Chern numbers},
    \href{http://link.aps.org/doi/10.1103/PhysRevB.86.241112}{Phys. Rev. B {\bf 86}, 241112(R) (2012)}.

\bibitem{ChernN}
Z. Liu, E.J. Bergholtz, H. Fan, and A.M. L\"auchli, {\em Fractional Chern Insulators in Topological Flat bands with Higher Chern Number},
    \href{http://link.aps.org/doi/10.1103/PhysRevLett.109.186805}{Phys. Rev. Lett. {\bf 109}, 186805 (2012)}.

\bibitem{ChernTwo}
Y.-F.~Wang, H.~Yao, C.-D.~Gong, and D. N. Sheng, {\em Fractional Quantum Hall Effect in Topological Flat Bands with Chern Number Two},
    \href{http://link.aps.org/doi/10.1103/PhysRevB.86.201101}{Phys. Rev. B {\bf 86}, 201101(R) (2012)}.

\bibitem{ChernN2}
A. Sterdyniak, C. Repellin, B. Andrei Bernevig, and N. Regnault, {\em Series of Abelian and non-Abelian states in $C>1$ fractional Chern insulators},
    \href{http://link.aps.org/doi/10.1103/PhysRevB.87.205137}{Phys. Rev. B {\bf 87}, 205137 (2013)}.

\bibitem{disloc} M. Barkeshli and X.-L. Qi, {\em Topological Nematic States and Non-Abelian Lattice Dislocations},
    \href{http://link.aps.org/doi/10.1103/PhysRevX.2.031013}{Phys. Rev. X {\bf 2}, 031013 (2012)}.

 \bibitem{andreas} A.M. L\"auchli, Z. Liu, E.J. Bergholtz, and R. Moessner, {\em Hierarchy of fractional Chern insulators and competing compressible states}, Phys. Rev. Lett., in press,
    \href{http://arxiv.org/abs/1207.6094}{arXiv:1207.6094}.

\bibitem{BHZ}B.A. Bernevig, T-L Hughes and S.-C. Zhang, {\em Quantum Spin Hall Effect and Topological Phase Transition in HgTe Quantum Wells}, \href{http://www.sciencemag.org/content/314/5806/1757.abstract}{Science {\bf 314}, 1757 (2006)}.
\\
\bibitem{topoexp} M. K\"onig, {\it et. al.}, 
{\em Quantum Spin Hall Insulator State in HgTe Quantum Wells}, \href{http://www.sciencemag.org/content/318/5851/766.abstract}{Science {\bf 318}, 766 (2007).}

\bibitem{Chernexp}C.-Z. Chang {\it et. al.}, 
{\em Experimental Observation of the Quantum Anomalous Hall Effect in a Magnetic Topological Insulator},
    \href{http://www.sciencemag.org/content/340/6129/167.abstract}{Science {\bf 340}, 6129 (2013)}.

\bibitem{hofstadtergraphene1} C. R. Dean {\it et. al.}, 
{\em HofstadterÕs butterfly and the fractal quantum Hall effect in moir\'e superlattices}, \href{http://dx.doi.org/10.1038/nature12186}{Nature {\bf 497}, 598 (2013)}.

\bibitem{hofstadtergraphene2}L. A. Ponomarenko {\it et. al.}, 
{\em Cloning of Dirac fermions in graphene superlattices},    \href{http://dx.doi.org/10.1038/nature12187}{Nature {\bf 497}, 594 (2013)}.

\bibitem{hofstadtercold} M. Atala, M. Lohse, J. T. Barreiro, B. Paredes, and I. Bloch, {\em Realization of the Hofstadter Hamiltonian with ultracold atoms in optical lattices}, \href{http://arxiv.org/abs/1308.0321}{arXiv:1308.0321}.


\bibitem{digital} D. Xiao, W. Zhu, Y. Ran, N. Nagaosa, and S. Okamoto,
{\em Interface engineering of quantum Hall effects in digital transition metal oxide heterostructures},
    \href{http://dx.doi.org/10.1038/ncomms1602}{Nat.~Commun. 2, 596 (2011)}.

\bibitem{ifw2}
J.W.F. Venderbos, S. Kourtis, J. van den Brink, and M. Daghofer,
{\em Fractional Quantum-Hall Liquid Spontaneously Generated by Strongly Correlated $t_{2g}$ Electrons},
    \href{http://link.aps.org/doi/10.1103/PhysRevLett.108.126405}{Phys. Rev. Lett. {\bf 108}, 126405 (2012)}.

\bibitem{DipolarTFB2}N. Y. Yao, A. V. Gorshkov, C. R. Laumann, A. M. L\"auchli, J. Ye, and M. D. Lukin,
{\em Realizing Fractional Chern Insulators with Dipolar Spins},
    \href{http://link.aps.org/doi/10.1103/PhysRevLett.110.185302}{Phys. Rev. Lett. {\bf 110}, 185302 (2013)}.

\bibitem{opticalflux}
N. R. Cooper and J. Dalibard, {\em Reaching Fractional Quantum Hall States with Optical Flux Lattices},
    \href{http://link.aps.org/doi/10.1103/PhysRevLett.110.185301}{Phys. Rev. Lett. {\bf 110}, 185301 (2013)}.

\bibitem{kolread}A. Kol, and N. Read, {\em Fractional quantum Hall effect in a periodic potential},
    \href{http://link.aps.org/doi/10.1103/PhysRevB.48.8890}{Phys. Rev. B {\bf 48}, 8890 (1993)}.

\bibitem{sorensen} A. S. S\/orensen, E. Demler, and M. D. Lukin, {\em Fractional Quantum Hall States of Atoms in Optical Lattices}, \href{http://link.aps.org/doi/10.1103/PhysRevLett.94.086803}{Phys. Rev. Lett. {\bf 94}, 086803 (2005)}.

\bibitem{palmer}
R. N. Palmer and D. Jaksch, {\em High-Field Fractional Quantum Hall Effect in Optical Lattices},
    \href{http://link.aps.org/doi/10.1103/PhysRevLett.96.180407}{Phys. Rev. Lett. {\bf 96}, 180407 (2006)}.

\bibitem{gunnar} G. Moller and N.R. Cooper, {\em Composite Fermion Theory for Bosonic Atoms in Optical Lattices}, \href{http://link.aps.org/doi/10.1103/PhysRevLett.103.105303}{Phys. Rev. Lett. {\bf 103}, 105303 (2009)}.

\bibitem{kalmeyerlaughlin}V. Kalmeyer and R. B. Laughlin, {\em Equivalence of the Resonating Valence Bond and Fractional Quantum Hall States}, \href{http://link.aps.org/doi/10.1103/PhysRevLett.59.2095}{Phys. Rev. Lett. {\bf 59}, 2095 (1987)}.

\bibitem{otherreview}
S. A. Parameswaran, R. Roy, and S. L. Sondhi, {\em Fractional Quantum Hall Physics in Topological Flat Bands},
    \href{http://arxiv.org/abs/1302.6606}{arXiv:1302.6606}.

\bibitem{nthroot} J.~McGreevy, B.~Swingle, K.-A.~Tran, {\em Wave functions for fractional Chern insulators}, \href{http://link.aps.org/doi/10.1103/PhysRevB.85.125105}{Phys. Rev. B {\bf 85}, 125105 (2012)}.

\bibitem{cherncf2}
G. Murthy, and R. Shankar, {\em  Hamiltonian theory of fractionally filled Chern bands},
    \href{http://link.aps.org/doi/10.1103/PhysRevB.86.195146}{Phys. Rev. B {\bf 86}, 195146 (2012)}.

\bibitem{hohenadler} M. Hohenadler and F. F. Assaad, {\em Correlation effects in two-dimensional topological insulators}, \href{http://stacks.iop.org/0953-8984/25/i=14/a=143201}{J. Phys.: Condens. Matter {\bf 25}, 143201 (2013)}.

\bibitem{Berry} M. V. Berry, {\em Quantal phase factors accompanying adiabatic changes}, \href{http://rspa.royalsocietypublishing.org/content/392/1802/45.abstract}{Proc. R. Soc. London A {\bf 392}, 45 (1984)}.

\bibitem{hallcond} T. Neupert, L. Santos, C. Chamon, and C.Mudry, {\em Elementary formula for the Hall conductivity of interacting systems},
    \href{http://link.aps.org/doi/10.1103/PhysRevB.86.165133}{Phys. Rev. B {\bf 86}, 165133 (2012)}.

\bibitem{topomott}
S. Raghu, X.-L. Qi, C. Honerkamp, and S.-C. Zhang, {\em Topological Mott Insulators},
    \href{http://link.aps.org/doi/10.1103/PhysRevLett.100.156401}{Phys. Rev. Lett. {\bf 100}, 156401 (2008)}.

\bibitem{nonab2}
Y.-F. Wang, H. Yao, Z.-C. Gu, C.-D. Gong, and D. N. Sheng,
{\em Non-Abelian Quantum Hall Effect in Topological Flat Bands},
    \href{http://link.aps.org/doi/10.1103/PhysRevLett.108.126805}{Phys. Rev. Lett. {\bf 108}, 126805 (2012)}.

\bibitem{ruby}X. Hu, M. Kargarian, and G. A. Fiete,
{\em Topological insulators and fractional quantum Hall effect on the ruby lattice},
    \href{http://link.aps.org/doi/10.1103/PhysRevB.84.155116}{Phys. Rev. B {\bf 84}, 155116 (2011)}.

\bibitem{ifw1}
J.W.F. Venderbos, M. Daghofer, and J. van den Brink,
{\em Narrowing of Topological Bands due to Electronic Orbital Degrees of Freedom},
    \href{http://link.aps.org/doi/10.1103/PhysRevLett.107.116401}{Phys. Rev. Lett. {\bf 107}, 116401 (2011)}.

\bibitem{coopermoessner}
N. R. Cooper and R. Moessner, {\em Designing Topological Bands in Reciprocal Space},
    \href{http://link.aps.org/doi/10.1103/PhysRevLett.109.215302}{Phys. Rev. Lett. {\bf 109}, 215302 (2012)}.

\bibitem{dario} D. Bercioux, N. Goldman, and  D. F. Urban, {\em Topology-induced phase transitions in quantum spin Hall lattices},    \href{http://link.aps.org/doi/10.1103/PhysRevA.83.023609}{Phys. Rev. A {\bf 83}, 023609 (2011)}.

\bibitem{disorder}S. Yang, K. Sun, and S. Das Sarma, {\em Quantum phases of disordered flatband lattice fractional quantum Hall systems}, \href{http://link.aps.org/doi/10.1103/PhysRevB.85.205124}{Phys. Rev. B {\bf 85}, 205124 (2012)}.

\bibitem{ifwlong} S. Kourtis, J. W. F. Venderbos, and M. Daghofer, {\em Fractional Chern insulator on a triangular lattice of strongly correlated $t_{2g}$ electrons},
    \href{http://link.aps.org/doi/10.1103/PhysRevB.86.235118}{Phys. Rev. B {\bf 86}, 235118 (2012)}.

\bibitem{qi}
X.-L.~Qi, {\em Generic Wavefunction Description of Fractional Quantum Anomalous Hall States and Fractional Topological Insulators},
    \href{http://link.aps.org/doi/10.1103/PhysRevLett.107.126803}{Phys. Rev. Lett. {\bf 107}, 126803 (2011)}.

\bibitem{BlochFCI}
Y.-L. Wu, N. Regnault, and B. A. Bernevig, {\em Bloch Model Wavefunctions and Pseudopotentials for All Fractional Chern Insulators},
    \href{http://link.aps.org/doi/10.1103/PhysRevLett.110.106802}{Phys. Rev. Lett. {\bf 110}, 106802 (2013)}.


\bibitem{bands1}
S.~A.~Parameswaran, R.~Roy, and S.~L.~Sondhi, {\em Fractional Chern Insulators and the W$_\infty$ Algebra},
    \href{http://link.aps.org/doi/10.1103/PhysRevB.85.241308}{Phys. Rev. B {\bf 85}, 241308(R) (2012)}.

\bibitem{bands2}
M.~O.~Goerbig,
{\em From Fractional Chern Insulators to a Fractional Quantum Spin Hall Effect},
    \href{http://stacks.iop.org/0953-8984/24/i=16/a=165503}{Eur.~Phys.~J.~B {\bf 85}, 15 (2012)}.

\bibitem{nonab1}
B.A. Bernevig and N. Regnault,
{\em Emergent many-body translational symmetries of Abelian and non-Abelian fractionally filled topological insulators},
    \href{http://link.aps.org/doi/10.1103/PhysRevB.85.075128}{Phys. Rev. B {\bf 85}, 075128 (2012)}.


\bibitem{roy} R. Roy, {\em Band geometry of fractional topological insulators},
    \href{http://arxiv.org/abs/1208.2055}{arXiv: 1208.2055}.

\bibitem{milica} E. Dobardzic, M.V. Milovanovic and N. Regnault, {\em On the geometrical description of fractional Chern insulators based on static structure factor calculations},
    \href{http://arxiv.org/abs/1303.7131}{arXiv:1303.7131}.

\bibitem{gmp} S. M. Girvin, A. H. MacDonald, and P. M. Platzman, {\em Magneto-roton theory of collective excitations in the fractional quantum Hall effect},
    \href{http://link.aps.org/doi/10.1103/PhysRevB.33.2481}{Phys. Rev. B {\bf 33}, 2481 (1986)}.


\bibitem{fubininoise}T. Neupert, C. Chamon and C. Mudry, {\em How to Measure the Quantum Geometry of Bloch Bands}, \href{http://link.aps.org/doi/10.1103/PhysRevB.87.245103}{Phys. Rev. B {\bf 87}, 245103 (2013)}.

\bibitem{Haldane85}
F.D.M. Haldane,
{\em Many-Particle Translational Symmetries of Two-Dimensional Electrons at Rational Landau-Level Filling},
    \href{http://link.aps.org/doi/10.1103/PhysRevLett.55.2095}{Phys. Rev. Lett. {\bf 55}, 2095 (1985)}.

\bibitem{bklong}E. J. Bergholtz and A. Karlhede, {\em Quantum Hall system in Tao-Thouless limit}, \href{http://link.aps.org/doi/10.1103/PhysRevB.77.155308}{Phys.~Rev.~B {\bf 77}, 155308 (2008)}; {\em 'One-dimensional' theory of the quantum Hall system},
    \href{http://stacks.iop.org/1742-5468/2006/i=04/a=L04001}{J.~Stat.~Mech. (2006) L04001}.

\bibitem{rootc}B. A. Bernevig and F. D. M. Haldane, {\em Model Fractional Quantum Hall States and Jack Polynomials}, \href{http://link.aps.org/doi/10.1103/PhysRevLett.100.246802}{Phys. Rev. Lett. {\bf 100}, 246802 (2008)}.

\bibitem{pattern}   X.-G. Wen and Z. Wang, {\em Topological properties of Abelian and non-Abelian quantum Hall states classified using patterns of zeros}, \href{http://link.aps.org/doi/10.1103/PhysRevB.78.155109}{Phys. Rev. B {\bf 78}, 155109 (2008)}.

\bibitem{pes}A. Sterdyniak, N. Regnault, and B. A. Bernevig, {\em Extracting Excitations from Model State Entanglement}, \href{http://link.aps.org/doi/10.1103/PhysRevLett.106.100405}{Phys. Rev. Lett. {\bf 106}, 100405 (2011)}.

\bibitem{nonab3}
Y.-L. Wu, B. A. Bernevig, and N. Regnault, {\em Zoology of Fractional Chern Insulators},
    \href{http://link.aps.org/doi/10.1103/PhysRevB.85.075116}{Phys.~Rev.~B {\bf 85}, 075116 (2012)}.

\bibitem{modularfci} W. Zhu, D. N. Sheng, and F. D. M. Haldane, {\em Minimal entangled states and modular matrix for fractional quantum Hall effect in topological flat bands}, \href{http://link.aps.org/doi/10.1103/PhysRevB.88.035122}{Phys. Rev. B {\bf 88}, 035122 (2013)}.

\bibitem{modularfci2} Y. Zhang and A. Vishwanath, {\em Establishing non-Abelian topological order in Gutzwiller-projected Chern insulators via entanglement entropy and modular S-matrix}, \href{http://link.aps.org/doi/10.1103/PhysRevB.87.161113}{Phys. Rev. B {\bf 87}, 161113(R) (2013).}

\bibitem{jain}
J. K. Jain,
{\em Composite-fermion approach for the fractional quantum Hall effect},
    \href{http://link.aps.org/doi/10.1103/PhysRevLett.63.199}{Phys.~Rev.~Lett. {\bf 63}, 199 (1989)};
J. K. Jain,  {\it Composite fermions}, (Cambridge University Press, 2007).

\bibitem{rr} N. Read and E. Rezayi, {\em Beyond paired quantum Hall states: Parafermions and incompressible states in the first excited Landau level}, \href{http://link.aps.org/doi/10.1103/PhysRevB.59.8084}{Phys.~Rev.~B {\bf 59}, 8084 (1999)}.

\bibitem{CompositeFCI}
T.~Liu, C.~Repellin, B.A.~Bernevig, and N.~Regnault,
{\em Fractional Chern Insulators beyond Laughlin states},
    \href{http://link.aps.org/doi/10.1103/PhysRevB.87.205136}{Phys. Rev. B {\bf 87}, 205136 (2013)}.

\bibitem{eliotzhao} Z. Liu, E. Kapit, and E.J. Bergholtz, in preparation.

\bibitem{dmrg}Z. Liu, D. L. Kovrizhin, and E. J. Bergholtz, {\em Bulk-edge correspondence in fractional Chern insulators}, \href{http://link.aps.org/doi/10.1103/PhysRevB.88.081106}{Phys. Rev. B {\bf 88}, 081106(R) (2013)}.

\bibitem{gaugefixing}
Y.-L.~Wu, N.~Regnault, and B.A.~Bernevig,
{\em Gauge-Fixed Wannier Wave-Functions for Fractional Topological Insulators},
    \href{http://link.aps.org/doi/10.1103/PhysRevB.86.085129}{Phys. Rev. B {\bf 86}, 085129 (2012)}.

\bibitem{LLlocal}E. I. Rashba,  L. E. Zhukov and A. L. Efros, {\em Orthogonal localized wave functions of an electron in a magnetic field},
    \href{http://link.aps.org/doi/10.1103/PhysRevB.55.5306}{Phys. Rev. B {\bf 55}, 5306 (1997)}.

\bibitem{hallabsence}D. J. Thouless, {\em Wannier functions for magnetic sub-bands},
    \href{http://stacks.iop.org/0022-3719/17/i=12/a=003}{J. Phys. C {\bf 17}, L325 (1984)}.

\bibitem{lam}P. K. Lam and S. M. Girvin, {\em Liquid-solid transition and the fractional quantum-Hall effect}, \href{http://link.aps.org/doi/10.1103/PhysRevB.30.473}{Phys. Rev. B {\bf 30}, 473 (1984)}.

\bibitem{wannier_review}N.~Marzari, A.~A.~Mostofi, J.~R.~Yates, I.~Souza, and D.~Vanderbilt, {\em Maximally localized Wannier functions: Theory and applications}, \href{http://link.aps.org/doi/10.1103/RevModPhys.84.1419}{Rev. Mod. Phys. {\bf 84}, 1419 (2012)}.

\bibitem{sv}A.~A.~Soluyanov and D.~Vanderbilt, {\em Wannier representation of $Z_2$ topological insulators}, \href{http://link.aps.org/doi/10.1103/PhysRevB.83.035108}{Phys. Rev. B {\bf 83}, 035108 (2011)}.

\bibitem{AdiabaticContinuity1}
T.~Scaffidi and G.~M\"oller,
{\em Adiabatic continuation of Fractional Chern Insulators to Fractional Quantum Hall States},
    \href{http://link.aps.org/doi/10.1103/PhysRevLett.109.246805}{Phys. Rev. Lett. {\bf 109}, 246805 (2012)}.

\bibitem{AdiabaticContinuity3}
Z.~Liu and E.J.~Bergholtz,
{\em From fractional Chern insulators to Abelian and non-Abelian fractional quantum Hall states: adiabatic continuity and orbital entanglement spectrum},
    \href{http://link.aps.org/doi/10.1103/PhysRevB.87.035306}{Phys. Rev. B {\bf 87}, 035306 (2013)}.

\bibitem{wanniersym}C.-M. Jian and X.-L. Qi, {\em Crystal-symmetry preserving Wannier states for fractional chern insulators}, \href{http://arxiv.org/abs/1303.1787}{arXiv:1303.1787}.

\bibitem{LiH}H.~Li and F.~D.~M.~Haldane, {\em Entanglement Spectrum as a Generalization of Entanglement Entropy: Identification of Topological Order in Non-Abelian Fractional Quantum Hall Effect States}, \href{http://link.aps.org/doi/10.1103/PhysRevLett.101.010504}{Phys. Rev. Lett. {\bf 101}, 010504, (2008)}.

\bibitem{AdiabaticContinuity2}
Y.-H.~Wu, J.K.~Jain, and K.~Sun,
{\em Adiabatic Continuity Between Hofstadter and Chern Insulators},
    \href{http://link.aps.org/doi/10.1103/PhysRevB.86.165129}{Phys. Rev. B {\bf 86}, 165129 (2012)}.

\bibitem{haldane83}
F.D.M. Haldane,
{\em Fractional Quantization of the Hall Effect: A Hierarchy of Incompressible Quantum Fluid States},
    \href{http://link.aps.org/doi/10.1103/PhysRevLett.51.605}{Phys. Rev. Lett. {\bf 51}, 605 (1983)}.

 \bibitem{wannierpp}C.H. Lee, R. Thomale, and X.-L. Qi,
{\em Pseudopotential Formalism for Fractional Chern Insulators},
    \href{http://link.aps.org/doi/10.1103/PhysRevB.88.035101}{Phys. Rev. B {\bf 88}, 035101 (2013)}.

\bibitem{haldanewfs}  F.D.M. Haldane and E. H. Rezayi, {\em Periodic Laughlin-Jastrow wave functions for the fractional quantized Hall effect}, \href{http://link.aps.org/doi/10.1103/PhysRevB.31.2529}{Phys. Rev. B {\bf 31}, 2529 (1985)}.

\bibitem{graphenepp}M. O. Goerbig, R. Moessner, and B.~Dou\c{c}ot, {\em Electron interactions in graphene in a strong magnetic field}, \href{http://link.aps.org/doi/10.1103/PhysRevB.74.161407}{Phys. Rev. B {\bf 74}, 161407(R) (2006)}.

\bibitem{pseudomulti}S. H. Simon, E. H. Rezayi, and N. R. Cooper, {\em Pseudopotentials for multiparticle interactions in the quantum Hall regime} \href{http://link.aps.org/doi/10.1103/PhysRevB.75.195306} {Phys. Rev. B {\bf 75}, 195306 (2007)}.

\bibitem{halperin84}
B.I. Halperin,
{\em Statistics of Quasiparticles and the Hierarchy of Fractional Quantized Hall States},
    \href{http://link.aps.org/doi/10.1103/PhysRevLett.52.1583}{Phys. Rev. Lett. {\bf 52}, 1583 (1984)}.

\bibitem{apf1}M. Levin, B. I. Halperin, and Bernd Rosenow, {\em Particle-Hole Symmetry and the Pfaffian State}, \href{http://link.aps.org/doi/10.1103/PhysRevLett.99.236806}{Phys. Rev. Lett. {\bf 99}, 236806 (2007)}.

\bibitem{apf2}S.-S. Lee, S. Ryu, C. Nayak, and M. P. A. Fisher, {\em Particle-Hole Symmetry and the $\nu=5/2$ Quantum Hall State}, \href{http://link.aps.org/doi/10.1103/PhysRevLett.99.236807}{Phys. Rev. Lett. {\bf 99}, 236807 (2007)}.


\bibitem{Grushin}
A.~G.~Grushin, T.~Neupert, C.~Chamon, and C.~Mudry, {\em Enhancing the stability of fractional Chern insulators against competing phases},
    \href{http://link.aps.org/doi/10.1103/PhysRevB.86.205125}{Phys. Rev. B {\bf 86}, 205125 (2012)}.

\bibitem{ThinTorusFCI}
B.~A.~Bernevig and N. Regnault,
{\em Thin-Torus Limit of Fractional Topological Insulators},
    \href{http://arxiv.org/abs/1204.5682}{arXiv:1204.5682}.

\bibitem{pan52} W. Pan {\it et. al.},
{\em Fractional Quantum Hall State at $\nu=5/2$ Landau Level Filling Factor},
    \href{http://link.aps.org/doi/10.1103/PhysRevLett.83.3530}{Phys. Rev. Lett. {\bf 83}, 3530 (1999)}.

\bibitem{masa} M. Udagawa, private communication (2012).

\bibitem{longrange}C.-M. Jian, Z.-C. Gu, and X.-L. Qi, {\em Momentum-space instantons and maximally localized flat-band topological Hamiltonians}, \href{http://dx.doi.org/10.1002/pssr.201206394}{Phys. Status Solidi (RRL) {\bf 7}, 154 (2013)}.

\bibitem{private} K. Sun and S. Yang, private communication.

\bibitem{zhao2}D. Wang, Z. Liu, J.-P. Cao, and H. Fan, {\em Generalized Hofstadter Model: Band Topology Transitions and Fractional Quantum Hall States}, \href{http://arxiv.org/abs/1304.7611}{arXiv:1304.7611}.

\bibitem{layla} L. Hormozi, G. Moller, and S. H. Simon, {\em Fractional Quantum Hall Effect of Lattice Bosons Near Commensurate Flux}, \href{http://link.aps.org/doi/10.1103/PhysRevLett.108.256809}{Phys. Rev. Lett. {\bf 108}, 256809 (2012)}.

\bibitem{joergsam}
J. Behrmann, S. Sanchez and E.J.~Bergholtz, in preparation.

\bibitem{111perovskites}S. Middey, D. Meyers, M. Kareev, E. J. Moon, B. A. Gray, X. Liu, J. W. Freeland, and J. Chakhalian, {\em Epitaxial growth of $(111)$-oriented $\textrm{LaAlO}_3/\textrm{LaNiO}_3$ ultra-thin superlattices},   \href{http://dx.doi.org/10.1063/1.4773375 }{Appl. Phys. Lett. {\bf 101}, 261602 (2012)}.

\bibitem{fiete}X. Hu, A. R\"uegg, and G. A. Fiete, {\em Topological phases in layered pyrochlore oxide thin films along the $[111]$ direction}, \href{http://link.aps.org/doi/10.1103/PhysRevB.86.235141}{Phys. Rev. B {\bf 86}, 235141 (2012)}.

\bibitem{organometallic}Z. Liu, Z.-F. Wang, J.-W. Mei, Y.-S. Wu and F. Liu, {\em Flat Chern Band in a Two-Dimensional Organometallic Framework}, \href{http://link.aps.org/doi/10.1103/PhysRevLett.110.106804}{Phys. Rev. Lett. {\bf 110}, 106804 (2013)}.

\bibitem{organometallic2}W. Li, Z. Liu, Y.-S. Wu, and Y.Chen, {\em Exotic Fractional Topological States in Two-Dimensional Organometallic Material}, \href{http://arxiv.org/abs/1308.1814}{arXiv:1308.1814}.

\bibitem{grapheneFCI}
P. Ghaemi, J. Cayssol, D. N. Sheng, and A. Vishwanath,
{\em Fractional Topological Phases and Broken Time-Reversal Symmetry in Strained Graphene},
    \href{http://link.aps.org/doi/10.1103/PhysRevLett.108.266801}{Phys. Rev. Lett. {\bf 108}, 266801 (2012)}.

\bibitem{artgauge}
J. Dalibard, F. Gerbier, G. Juzeliunas, and P. Ohberg, {\em Colloquium: Artificial gauge potentials for neutral atoms}, \href{http://link.aps.org/doi/10.1103/RevModPhys.83.1523}{Rev. Mod. Phys. {\bf 83}, 1523 (2011)}.

\bibitem{viewpoint2} M. Daghofer and M. Haque, {\em Toward Fractional Quantum Hall Physics with Cold Atoms},
\href{http://link.aps.org/doi/10.1103/Physics.6.49}{Physics {\bf 6}, 49 (2013)}.

\bibitem{DipolarTFB}
N.Y.~Yao, C.R.~Laumann, A.V.~Gorshkov, S.D.~Bennett, E.~Demler, P.~Zoller, and M.D.~Lukin,
{\em Topological Flat Bands from Dipolar Spin Systems}, \href{http://link.aps.org/doi/10.1103/PhysRevLett.109.266804}{Phys. Rev. Lett. {\bf 109}, 266804 (2012)}.

\bibitem{taoshi}T. Shi and J. I. Cirac, {\em Topological phenomena in trapped-ion systems}, \href{http://link.aps.org/doi/10.1103/PhysRevA.87.013606}{Phys. Rev. A {\bf 87}, 013606 (2013)}.

\bibitem{tfl}F.~D.~M.~Haldane, {\em Berry Curvature on the Fermi Surface: Anomalous Hall Effect as a Topological Fermi-Liquid Property}, \href{http://link.aps.org/doi/10.1103/PhysRevLett.93.206602}{Phys. Rev. Lett. {\bf 93}, 206602 (2004)}.

\bibitem{tfl2} K.~Sun and E.~Fradkin, {\em Time-reversal symmetry breaking and spontaneous anomalous Hall effect in Fermi fluids}, \href{http://link.aps.org/doi/10.1103/PhysRevB.78.245122}{Phys. Rev. B {\bf 78}, 245122 (2008)}.

\bibitem{hlr} B.I. Halperin, P.A. Lee, and N. Read, {\em Theory of the half-filled Landau level}, \href{http://link.aps.org/doi/10.1103/PhysRevB.47.7312}{Phys. Rev. B {\bf 47}, 7312 (1993)}.

\bibitem{cdw+fci}
S. Kourtis and M. Daghofer, {\em Combined topological and Landau order from strong correlations in Chern bands}, \href{http://arxiv.org/abs/1305.6948}{arXiv:1305.6948}.

\bibitem{haldanegeometry} F. D. M. Haldane, {\em Geometrical Description of the Fractional Quantum Hall Effect}, \href{http://link.aps.org/doi/10.1103/PhysRevLett.107.116801}{Phys. Rev. Lett. {\bf 107}, 116801 (2011)}.

\bibitem{ttmps}M.~Nakamura, Z.-Y.~Wang, and E.~J.~Bergholtz, {\em Exactly Solvable Fermion Chain Describing a $\nu=1/3$ Fractional Quantum Hall State},
    \href{http://link.aps.org/doi/10.1103/PhysRevLett.109.016401}{Phys. Rev. Lett. {\bf 109}, 016401 (2012)}.

 \bibitem{pairingproblem}   G. Ortiz, Z. Nussinov, J. Dukelsky, and A. Seidel, {\em Repulsive Interactions in Quantum Hall Systems as a Pairing Problem}, \href{http://arxiv.org/abs/1306.3268}{arXiv:1306.3268}.

\bibitem{mpsfqh}M.~P.~Zaletel and R.~S.~K.~Mong, {\em Exact matrix product states for quantum hall wave functions}, \href{http://link.aps.org/doi/10.1103/PhysRevB.86.245305}{Phys. Rev. B {\bf 86} 245305 (2012)}.

\bibitem{mpsfqh2}B.~Estienne, Z.~Papic, N.~Regnault, and B.~A.~Bernevig, {\em Matrix product states for trial quantum Hall states}, \href{http://link.aps.org/doi/10.1103/PhysRevB.87.161112}{Phys. Rev. B {\bf 87}, 161112(R) (2013)}.

\bibitem{mpsfqh3}M.~P.~Zaletel, R.~S.~K.~Mong, and F.~Pollmann, {\em Topological Characterization of Fractional Quantum Hall Ground States from
Microscopic Hamiltonians}, \href{http://link.aps.org/doi/10.1103/PhysRevLett.110.236801}{Phys. Rev. Lett. {\bf 110}, 236801 (2013)}.

\bibitem{cincio} L.~Cincio and G.~Vidal, {\em Characterizing Topological Order by Studying the Ground States on an Infinite Cylinder}, \href{http://link.aps.org/doi/10.1103/PhysRevLett.110.067208}{Phys. Rev. Lett. {\bf 110}, 067208 (2013)}.

\bibitem{balents}H.-C.~Jiang, Z.~Wang and L.~Balents, {\em Identifying topological order by entanglement entropy}, \href{http://dx.doi.org/10.1038/nphys2465}{Nature Physics {\bf 8}, 902 (2012)}.

\bibitem{arealaw}J.~Eisert, M.~Cramer, and M.~B.~Plenio, {\em Area laws for the entanglement entropy}, \href{http://link.aps.org/doi/10.1103/RevModPhys.82.277}{Rev. Mod. Phys. {\bf 82}, 277 (2010)}.

\bibitem{tnreview}F.~Verstraete, V.~Murg, and J.~I.~Cirac, {\em Matrix product states, projected entangled pair states, and variational renormalization group methods for quantum spin systems}, \href{http://www.tandfonline.com/doi/abs/10.1080/14789940801912366}{Adv. Phys. {\bf 57}, 143 (2008)}.

\bibitem{tnreview2}N.~Schuch, J.~I.~Cirac, and D.~Perez-Garcia, {\em PEPS as ground states: Degeneracy and topology}, \href{http://dx.doi.org/10.1016/j.aop.2010.05.008}{Ann. Phys. {\bf 325}, 2153 (2010)}.

\bibitem{tensorchiral} J. Dubail and N. Read, {\em Tensor network trial states for chiral topological phases in two dimensions}, \href{http://arxiv.org/abs/1307.7726}{arXiv:1307.7726}.

\bibitem{tensorchiral2} T.B. Wahl, H.-H. Tu, N. Schuch, and J.I. Cirac, {\em Projected entangled-pair states can describe chiral topological states}, \href{http://arxiv.org/abs/1308.0316}{arXiv:1308.0316}.

\bibitem{beri} B. Beri and N. R. Cooper, {\em Local Tensor Network for Strongly Correlated Projective States}, \href{http://link.aps.org/doi/10.1103/PhysRevLett.106.156401}{Phys. Rev. Lett. {\bf 106}, 156401 (2011)}.


\bibitem{jan} J. C. Budich and E. Ardonne, {\em Fractional topological phase in one-dimensional flat bands with nontrivial topology}, \href{http://link.aps.org/doi/10.1103/PhysRevB.88.035139}{Phys. Rev. B {\bf 88}, 035139 (2013)}.

\bibitem{kevin}K. A. Madsen, E. J. Bergholtz and P. W. Brouwer {\em Topological equivalence of crystal and quasicrystal band structures}, \href{http://arxiv.org/abs/1307.2577}{arXiv:1307.2577}.

\bibitem{thoulesspump}D. J. Thouless, {\em Quantization of particle transport}, \href{http://link.aps.org/doi/10.1103/PhysRevB.27.6083}{Phys. Rev. B {\bf 27}, 6083 (1983)}.

\bibitem{seidel} A. Seidel, {\em S-Duality Constraints on $1D$ Patterns Associated with Fractional Quantum Hall States},  \href{http://link.aps.org/doi/10.1103/PhysRevLett.105.026802}{Phys. Rev. Lett. {\bf 105}, 026802 (2010)}.

\bibitem{FTI}M. Levin and A. Stern, {\em Fractional topological insulators}, \href{http://link.aps.org/doi/10.1103/PhysRevLett.103.196803}{Phys.~Rev.~Lett. {\bf 103}, 196803 (2009)}.

\bibitem{qsh}B. A. Bernevig and S.-C. Zhang, {\em Quantum Spin Hall Effect},  \href{http://link.aps.org/doi/10.1103/PhysRevLett.96.106802}{Phys. Rev. Lett. {\bf 96}, 106802 (2006)}.

\bibitem{neupertfti}T.~Neupert, L.~Santos, S.~Ryu, C.~Chamon, and C.~Mudry, {\em Fractional topological liquids with time-reversal symmetry and their lattice realization}, \href{http://link.aps.org/doi/10.1103/PhysRevB.84.165107}{Phys. Rev. B {\bf 84}, 165107 (2011)}.

\bibitem{neupertfti2} L.~Santos, T.~Neupert, S.~Ryu, C.~Chamon, and C.~Mudry, {\em Time-reversal symmetric hierarchy of fractional incompressible liquids}, \href{http://link.aps.org/doi/10.1103/PhysRevB.84.165138}{Phys. Rev. B {\bf 84}, 165138 (2011)}.

\bibitem{peng}P. Ye and X.-G. Wen, {\em Projective construction of two-dimensional symmetry-protected topological phases with $U(1)$, $SO(3)$, or $SU(2)$ symmetries}, \href{http://link.aps.org/doi/10.1103/PhysRevB.87.195128}{Phys. Rev. B {\bf 87}, 195128 (2013)}.

\bibitem{flatsurface}M. Sitte, A. Rosch and L. Fritz, {\em Interaction effects on almost flat surface bands in topological insulators},     \href{http://arxiv.org/abs/1305.1788}{arXiv:1305.1788}.

\bibitem{surfaceFQH}C. Wang, A. C. Potter and T. Senthil, {\em Gapped Symmetry Preserving Surface-State for the Electron Topological Insulator},
    \href{http://arxiv.org/abs/1306.3223}{arXiv:1306.3223}.

\bibitem{surfaceFQH2}M. A. Metlitski, C. L. Kane and M. P. A. Fisher, {\em A symmetry-respecting topologically-ordered surface phase of $3D$ electron topological insulators},
    \href{http://arxiv.org/abs/1306.3286}{arXiv:1306.3286}.

\bibitem{surfaceFQH3}P. Bonderson, C. Nayak and X.-L. Qi, {\em A Time-Reversal Invariant Topological Phase at the Surface of a $3D$ Topological Insulator},
    \href{http://arxiv.org/abs/1306.3230}{arXiv:1306.3230}.

\bibitem{surfaceFQH4}X. Chen, L. Fidkowski and A. Vishwanath, {\em Symmetry Enforced Non-Abelian Topological Order at the Surface of a Topological Insulator}
    \href{http://arxiv.org/abs/1306.3250}{arXiv:1306.3250}.

\bibitem{3dintclass}C. Wang, A. C. Potter and T. Senthil, {\em Classification of interacting electronic topological insulators in three dimensions}
    \href{http://arxiv.org/abs/1306.3238}{arXiv:1306.3238}.

\bibitem{3dFTI} J. Maciejko, X.-L. Qi, A. Karch, and S.-C. Zhang, {\em Fractional Topological Insulators in Three Dimensions},    \href{http://link.aps.org/doi/10.1103/PhysRevLett.105.246809}{Phys. Rev. Lett. {\bf 105}, 246809 (2010)}.

\bibitem{3dFTI2}B. Swingle, M. Barkeshli, J. McGreevy, and T. Senthil, {\em Correlated Topological Insulators and the Fractional Magnetoelectric Effect}, \href{http://link.aps.org/doi/10.1103/PhysRevB.83.195139}{Phys. Rev. B {\bf 83}, 195139 (2011)}.

\bibitem{3dflatbands}  C. Weeks and M. Franz, {\em Flat bands with nontrivial topology in three dimensions},
\href{http://link.aps.org/doi/10.1103/PhysRevB.85.041104}{Phys. Rev. B {\bf 85}, 041104(R) (2012)}.


\end{thebibliography}
\end{document}